\documentclass[12pt]{iopart}
\usepackage{amssymb}
\usepackage{epsfig}
\usepackage{graphicx}
\begin{document}
\title[Nonlinear modes for the Gross-Pitaevskii equation]{Nonlinear modes for the
Gross-Pitaevskii equation -- demonstrative
computation approach}

\author{G L Alfimov and D A Zezyulin}

\address{Moscow Institute of Electronic Engineering,
    Zelenograd, Moscow, 124498, Russia}
\ead{galfimov@yahoo.com, d.zezyulin@gmail.com}

\begin{abstract}
A method for the study of steady-state nonlinear modes for
Gross-Pitaevskii equation (GPE) is described. It is based on exact
statement about coding of the steady-state solutions of GPE which
vanish as $x\to+\infty$ by reals. This allows to fulfill {\it
demonstrative computation} of nonlinear modes of GPE i.e. the
computation which allows to guarantee that {\it all} nonlinear modes
within a given range of parameters have been found. The method has been
applied to GPE with quadratic and double-well potential, for both,
repulsive and attractive nonlinearities. The bifurcation diagrams of
nonlinear modes in these cases are represented. The stability of these
modes has been discussed.
\end{abstract}
 \submitto{\NL} \maketitle

\maketitle

\section{Introduction}
The Gross-Pitaevskii equation (GPE) is one of the basic models in the
theory of Bose-Einstein condensate (BEC)\cite{Pitaev}. In dimensionless
form this equation reads
\begin{equation}
i \Psi_t=\Delta\Psi-V({\bf x})\Psi+\sigma \Psi |\Psi|^2
,\label{GPEq}
\end{equation}
In BEC context GPE describes an atomic cloud in the mean-field
approximation. The function $\Psi(t,{\bf x})$ corresponds to
macroscopic wave function, $\Delta$ is the Laplace operator and
$V({\bf x})$ is a trap potential for the condensate to be
confined. The value of $\sigma$ governs the sign of nonlinearity:
the case $\sigma=1$ corresponds to  attractive interatomic
interactions and $\sigma=-1$ corresponds to repulsive interatomic
interactions. Both the cases $\sigma=\pm 1$ are of interest for
physical applications \cite{Science1995,PRL1995}. Eq.(\ref{GPEq})
conserves the quantities
\begin{eqnarray}
&&E=\frac 12\int \left(|\nabla \Psi|^2-V({\bf
x})|\Psi|^2-\frac\sigma 2|\Psi|^4\right)~d{\bf x}\label{Energy}\\
&&N=\int |\Psi|^2~d{\bf x} \label{N}
\end{eqnarray}
which have the sense of Hamiltonian and number of particles
correspondingly. An important class of solutions of GPE are
stationary nonlinear modes defined as
\begin{equation}
\Psi(t,{\bf x})=e^{i\omega t}\psi({\bf x})\label{Anzats}
\end{equation}
\begin{equation}
\psi({\bf x})\to 0\quad  \mbox{as}\quad |{\bf x}|\to
\infty\label{Local}
\end{equation}
In this case $\psi({\bf x})$ satisfies the equation
\begin{equation}
\Delta\psi+(\omega-V({\bf x}))\psi+\sigma \psi |\psi|^2
=0,\label{StatGPEq}
\end{equation}
The parameter $\omega$ in terms of BEC corresponds to the chemical
potential. The {\it ground state} solutions of
(\ref{Local})-(\ref{StatGPEq}) (i.e. node-free solutions which minimize
energy functional for Eq.(\ref{GPEq})) are of primary importance for
BEC applications \cite{EdwBur1995,Rupr1995,DalfString1996}. Apart from
them some {\it non-ground} (i.e. excited) nonlinear modes also have
been reported (see e.g.
\cite{Yukalov1997,Kivshar2001,MalomLasPhys2002,AgPres2002,Gammal2006})
as well as physical mechanisms for their generation \cite{Yukalov2002}.
Except some specific examples of the potential $V({\bf x})$ when exact
solutions of (\ref{Local})-(\ref{StatGPEq}) are available (see e.g.
\cite{Square1}) in all the cases the nonlinear states have been found
numerically.

Typically, numerical calculation of nonlinear modes needs some initial
guess for subsequent iterative procedure. In order to get  {\it a
priori} information about the shape of nonlinear states various
approaches have been used. In \cite{Kivshar2001} the branches of
nonlinear modes for the case $V({\bf x})=|{\bf x}|^2$ have been found
by continuation from the limit $\psi\to 0$ where the nonlinear term can
be neglected
\begin{eqnarray}
\Delta\psi+(\omega-|{\bf x}|^2)\psi =0.\label{LinGPEq}
\end{eqnarray}
Therefore, the shapes of these states can be described by the set
of Gauss-Hermite functions.  In 1D case and for $\sigma=1$ it has
been proved rigorously \cite{GaussTail} that there are infinitely
many branches of nonlinear modes which bifurcate from zero
solution of (\ref{StatGPEq}) at discrete infinite set of values of
$\omega$ corresponding to eigenvalues of linear problem
(\ref{LinGPEq}). Following \cite{MalomLasPhys2002,AgPres2002}, it
is quite natural to call these modes {\it the modes with linear
counterpart}. It has been shown in
\cite{MalomLasPhys2002,AgPres2002} that for some potentials
$V({\bf x})$ nonlinear modes without linear counterpart also can
exist. In the case of multi-well potentials  these states have
been found by continuation from asymptotical  limit of large
nonlinearity \cite{MalomLasPhys2002,AgPres2002}.

However, the following question remains: can one list {\it all}
nonlinear states corresponding to given potential $V({\bf x})$ with
$\omega$ lying within some range? To best of our knowledge, up to the
moment there is no answer to this question even in the most explored
1D-case of quadratic potential $V(x)=x^2$. The aim of this paper is to
make a step toward answering this question. Specifically, for
$\sigma=-1$ we explain how to arrange numerical calculation  in such a
way that {\it all} nonlinear modes described by Eq.(\ref{StatGPEq}) for
$\omega$ lying within a given range would be found. In the case
$\sigma=1$ the method gives a picture which is complete, as we believe,
but this fact has not been proved. We illustrate our approach on 1D
case of Eq.(\ref{StatGPEq}) with quadratic and double-well potentials
(the applications to radial 2D and 3D cases are straightforward). The
base for this approach is given by {\it exact} statement on one-to-one
correspondence between the solutions of Eq.(\ref{StatGPEq}) which tends
to zero as $x\to+\infty$ and linear coefficient $C_+$ which determines
their asymptotics in this limit.

So, in this paper we consider 1D version of Eq.(\ref{StatGPEq}),
specifically
\begin{equation}
\psi''+(\omega-V(x))\psi+\sigma \psi^3=0 \label{MainEq}
\end{equation}
where prime denotes differentiation with respect to $x$. We assume that
the potential $V(x)$ is {\it even}, $V(x)=V(-x)$ and grows when
$|x|\to\infty$. The attention is focused on two particular examples of
the potential,
\begin{eqnarray}
&&V(x)=x^2,\quad \mbox{(quadratic potential)}\label{TrapHarm}\\
&&V(x)=\alpha x^4+\beta x^2,\quad \alpha>0,\quad
\beta<0\quad\mbox{(double-well potential)}\label{TrapDW}
\end{eqnarray}
The solutions of  Eq.(\ref{MainEq}) are supposed to be real and satisfy
the localization condition
\begin{equation}
\lim_{x\to\pm\infty}\psi(x)=0\label{LocalCond}
\end{equation}

The paper is organized as follows. In Section \ref{back} we describe
the method which we use. In Section \ref{Numeric} we apply this method
to Eq.(\ref{MainEq}) with quadratic and double-well potentials. Section
\ref{Concl} contains summary and discussion. Some auxiliary statements
and technical details can be found in Appendices A-C.

\section{The description of the approach}\label{back}

\subsection{The idea of the approach}\label{back_idea}

The approach which we propose is an extension of one used in
\cite{Rupr1995,Gammal1999} for numerical construction of nonlinear
modes for Eq.(\ref{StatGPEq}). Let us denote by $S_+$ the set of all
solutions of Eq.(\ref{MainEq}) which satisfy the condition
\begin{equation}
\lim_{x\to+\infty} \psi(x) =0\label{Class0}
\end{equation}
When $x\to+\infty$ the nonlinear term in Eq.(\ref{MainEq}) can be
neglected  and the asymptotics of solution from $S_+$ can be found from
linearized equation
\begin{equation}
\psi''+(\omega-V(x))\psi=0 \label{LinMainEq}
\end{equation}
The theory of Eq.(\ref{LinMainEq}) (see e.g.
\cite{BerShub,Fedoryuk}) says that for a wide class of potentials
$V(x)$ it has {\it unique} (up to constant factor) solution
$\tilde{\psi}(x)$ such that $\tilde{\psi}(x)\to 0$ when
$x\to+\infty$. If $V(x)>\omega$ for $x>x_0$ for some $x_0$ large
enough (and fixed) then the asymptotics of this specific solution
is described by the formula  \cite{Fedoryuk}
\begin{equation}
\tilde{\psi}(x)= \frac
{e^{-\int\limits_{x_0}^x\sqrt{V(t)-\omega}~dt}}{\sqrt[4]{V(x)-\omega}}\left(1
+o(1)\right), \quad x\to+\infty\label{LinGenAsympt}
\end{equation}
Then the solution of Eq.(\ref{MainEq}), $\psi(x)$, which satisfies
(\ref{Class0}) has the asymptotics
\begin{equation}
\psi(x)= \frac
{e^{-\int\limits_{x_0}^x\sqrt{V(t)-\omega}~dt}}{\sqrt[4]{V(x)-\omega}}\left(C_+
+o(1)\right), \quad x\to+\infty\label{GenAsympt+}
\end{equation}
Here $C_+$ is some constant. The next step is to assume that there is
one-to-one correspondence between the  solutions from $S_+$ and the
constants $C_+\in {\bf R}$ which comes from formula (\ref{GenAsympt+})
(this fact will be rigorously justified below). This implies that the
solutions $S_+$ constitute one parameter set with free parameter
$C_+\in {\bf R}$. Let us denote a solution of Eq.(\ref{MainEq}) which
satisfies the asymptotical relation (\ref{GenAsympt+}) by
$\psi_{+}(x,C_+)$. By construction, the solutions $\psi_{+}(x;C_+)$
tend to zero as $x\to+\infty$; however, generically, they do not tend
to zero as $x\to -\infty$.

The same reasoning holds for the solutions of Eq.(\ref{MainEq})
which tend to zero as $x\to -\infty$ (the set of these solutions
will be denoted by $S_-$).The asymptotics of these solutions is of
the form
\begin{equation}
\psi(x)= \frac
{e^{\int\limits^{-x_0}_x\sqrt{V(t)-\omega}~dt}}{\sqrt[4]{V(x)-\omega}}\left(C_-
+o(1)\right), \quad x\to-\infty\label{GenAsympt-}
\end{equation}
where  $C_-$ is some constant. Assume that $S_-$ is one-parameter
set with free parameter $C_-$. Let us denote a solution of
Eq.(\ref{MainEq}) which satisfies the asymptotical relation
(\ref{GenAsympt-}) by $\psi_{-}(x,C_-)$. Evidently, a localized
solution $\psi(x)$ which satisfies (\ref{LocalCond}) must belong
to both sets, $S_+$ and $S_-$, and in the point $x=0$ the
relations must be valid
\begin{equation}
\psi(0)=\psi_{-}(0,C_-)=\psi_{+}(0,C_+);\quad
\psi'(0)=\psi_{-}'(0,C_-)=\psi_{+}'(0,C_+); \label{C+C-}
\end{equation}
In fact, (\ref{C+C-}) is a system of equations to determine
unknown constants $C_-$ and $C_+$ and, consequently, a solution
$\psi(x)$. In practice, the values $\psi_{+}(0;C_+)$ and
$\psi_{+}'(0,C_+)$ can be calculated numerically (using, for
example, Runge-Kutta method) by solving the initial value problem
for Eq.(\ref{MainEq}) starting  from some large enough value
$x=x_\infty$ where  the values $\psi_{+}(x_\infty;C_+)$,
$\psi'_{+}(x_\infty;C_+)$ can be approximated by means of
asymptotical formula (\ref{GenAsympt+}). In the same way the
values $\psi_{-}(0;C_-)$, $\psi'_{-}(0;C_-)$ can be calculated.

\subsection{Mathematical basis}\label{back_basis}

The heuristical arguments given in paragraph \ref{back_idea} can be
justified by exact statement which describes the set of solutions of
Eq.(\ref{MainEq}) which tend to zero as $x\to+\infty$. This can be done
using ``asymptotic integration'' of Eq.(\ref{MainEq}) (see e.g.
\cite{Hartman}).   The following statement is valid:

\noindent{\it \underline{Theorem 1.} Let there exists some $x_0$ such
that
\medskip

(i) $V(x)-\omega\geq\varepsilon>0$ for $x>x_0$;
\medskip

(ii) $V(x)\in C^2(x_0,+\infty)$;
\medskip

(iii) $\int_{x_0}^\infty
\frac{|V''(x)|}{(V(x)-\omega)^{3/2}}~dx<\infty$
\medskip

Let $\psi_1(x)$ be a fixed solution of linear equation
(\ref{LinMainEq}) such that $\psi_1(x)\to 0$ as $x\to+\infty$ which has
asymptotics (\ref{LinGenAsympt}). Then
\medskip

(a) for any solution $\psi(x)\in S_+$ there exists a limit
\begin{equation}
\lim_{x\to +\infty}\frac{\psi(x)}{\psi_1(x)}=C_+<\infty \label{Ass1}
\end{equation}

(b) for any $C_+\in{\bf R}$  there exists an {\em unique} solution
$\psi_+(x;C_+)\in S_+$ of Eq.(\ref{MainEq}) such that
\begin{equation}
\lim_{x\to +\infty}\frac{\psi_+(x,C_+)}{\psi_1(x)}=C_+ \label{Ass2}
\end{equation}

(c) if  $\psi_+(x;C_+)$ exists in some neighbourhood $U$ of a
point $(\tilde x;\tilde{C}_+)$ then in $U$ there also exists
partial derivative $\frac{\partial\psi_+(x;C_+)}{\partial C_+}$
which is continuous with respect to $C_+$.}
\bigskip

Theorem 1 establishes one-to-one correspondence between $S_+$ and ${\bf
R}$; specifically, it says that the set $S_+$ can be parametrized by
real parameter $C_+\in{\bf R}$. The point (a) of Theorem 1 has been
used for the search of localized solutions by means of shooting method
(see e.g. \cite{Gammal1999}). We make use of the point (b) to construct
{\it general picture} of branches of solutions of Eq.(\ref{MainEq}).
The point (c) will be used for proving Lemma 1 in Appendix C.

Theorem 1 reflects the well-known relation between the behavior of
linear dynamical system and its nonlinear perturbation in a vicinity of
its zero solution.  In spite of bulk of literature devoted to this
subject (see, for instance, \cite{Rodrigues} and references therein) we
failed to find this statement {\it exactly} in the form we need.
Therefore we adduce a sketch of proof of Theorem 1 in Appendix B in a
little more general form. Evidently, analogous statement can be made
for the solutions of Eq.(\ref{MainEq}) which tend to zero as
$x\to-\infty$.

\subsection{Practical implementation of the method and visualization of the results}\label{back_visio}

\subsubsection{$(\psi(0),\psi'(0))$-pictures.}\label{psi_pic}

Let the potential $V(x)$ be {\it even}, $V(x)=V(-x)$. Let $\omega$
be fixed and  $x=x_{\infty}$ is some value large enough where the
solution of (\ref{MainEq}) can be approximated well by the
asymptotics (\ref{GenAsympt+}). Assume that the numerical
continuation of the solutions $\psi_{+}(x;C_+)$ from
$x=x_{\infty}$ to $x=0$ can be fulfilled for each $C_+$ in some
interval $C_+\in[-\tilde{C}_+;\tilde{C}_+]$ (using, for instance,
Runge-Kutta method). The result can be represented as a segment of
curve
\begin{eqnarray*}
\gamma_+:\{(\psi_+(0;C_+);\psi'_+(0;C_+)),C_+\in[-\tilde{C}_+;\tilde{C}_+]\}
\end{eqnarray*}
on the plane $(\psi;\psi')$. Since $C_+=0$ corresponds to zero
solution, this curve passes through the point $(0;0)$ of this
plane. Moreover, due to the fact that the equation (\ref{MainEq})
admits the symmetry $\psi\to-\psi$, this segment of the curve is
centrosymmetric with respect to the origin. In practice this means
that the calculation should be fulfilled for $[0;\tilde{C}_+]$
only. In analogous way, let us consider the segment of curve
\begin{eqnarray*}
\gamma_-:\{(\psi_-(0;C_-);\psi'_-(0;C_-)),C_-\in[-\tilde{C}_-;\tilde{C}_-]\}
\end{eqnarray*}
constructed by means of numerical continuation of the solution
$\psi_{-}(x;C_-)$ from the value $x=-x_{\infty}$ until the value
$x=0$. Since equation (\ref{MainEq}) is  invariant with respect to
the change $x\to -x$, if $\tilde{C}_-=\tilde{C}_+$ the curve
$\gamma_-$ is symmetric about the $\psi'$ axis to the curve
$\gamma_+$. Then, each point of intersection of $\gamma_-$ and
$\gamma_+$ corresponds to a solution of system (\ref{C+C-}) and,
consequently, to a solution of Eq.(\ref{MainEq}) with the
condition (\ref{LocalCond}) (see Fig.\ref{Spirals}).

Due to the symmetries of the curves $\gamma_-$ and $\gamma_+$, the
points of intersection of $\gamma_+$ with the axes $\psi$ and
$\psi'$ are also
 points of intersection of $\gamma_+$ with the curve $\gamma_-$.
These points correspond to specific  localized solutions of
Eq.(\ref{MainEq}); specifically, the points of intersection of
$\gamma_+$ with the axis $\psi$ correspond to {\it even} localized
solutions of Eq.(\ref{MainEq}) whereas the points of intersection of
$\gamma_+$ with the axis $\psi'$ correspond to {\it odd} localized
solutions of this equation. The points of intersection of $\gamma_-$
and $\gamma_+$ which do not belong to the axes correspond to
non-symmetric  localized solutions of Eq.(\ref{MainEq}).

\subsubsection{$(C_+,\omega)$-diagrams.}\label{C-om_pic}

Let us consider now   branches of localized solutions of
Eq.(\ref{MainEq}) which appear when the parameter $\omega$ varies. Due
to Theorem 1, if $\omega$ is fixed, each of the localized solutions of
Eq.(\ref{MainEq}) can be unambiguously defined by the value $C_+$
determined by its asymptotics. Therefore, it is convenient to represent
the  branches of localized solutions on the plane $(C_+,\omega)$
marking for each $\omega$ those values $C_+$ for which localized
solutions exist. In such a way any branch of solutions corresponds to a
curve $\Gamma$ on the plane $(C_+,\omega)$. An advantage of this
graphical representation (in comparison, for instance, with more
traditional $(N,\omega)$ representation where $N$ is given by
(\ref{N})) is that {\it each point of the plane $(C_+,\omega)$
corresponds to at most one localized solution } of Eq.(\ref{MainEq}).
This means, for instance, that if two curves, corresponding to two
branches have a common point, then there exists a localized solution
which belongs to both branches. This is not the case for
$(N,\omega)$-diagrams.

Consider now any segment of curve $\Gamma$ which has no intersection
with the axis $C_+=0$. Simple reasoning shows that {\it the number of
zeros of solutions $\psi(x)$ is constant along this segment of
$\Gamma$}. Indeed, consider a continuous deformation of solution
$\psi(x)$ which occurs when passing along this segment. A zero $x=x_0$
of $\psi(x)$ can disappear by (i) merging with one or several
neighboring zeros; (ii) by merging with $x=\pm\infty$ (``goes to
infinity''). The case (i) implies that the branch includes a solution
with $\psi(x_0)=\psi'(x_0)=0$ which is zero solution. Consider the case
(ii) i.e. assume that zero $x=x_0$ merges with $x=+\infty$. Then for
some solution of the branch the parameter $C_+$ is zero which implies
that the solution is identically zero. Similar arguments allow to
discard also the case when  (i) and (ii) hold simultaneously.

\subsection{Some additional statements for the case of repulsive nonlinearity}\label{rep_non_stat}

In the case of repulsive nonlinearity Eq.(\ref{MainEq}) reads
\begin{equation}
\psi''+(\omega-V(x))\psi- \psi^3=0 \label{MainEqRep}
\end{equation}
It turns out that ``most part'' of solutions of this equation {\it are
singular} i.e. they tend to $\pm\infty$ as $x$ approaches some finite
value. This fact allows to fulfill more complete analysis of localized
solutions for Eq.(\ref{MainEqRep}). It can be based on the following
statement:

{\it \underline{Lemma 1} Let there exists a solution
$\psi_0(x)=\psi_+(x,C^0_+)\in S_+$, $C^0_+>0$ of equation
(\ref{MainEqRep}) such that:
\bigskip

(i) $\psi_0(x)$ is defined on $(x_0;+\infty)$ where $x_0>0$ and
\begin{eqnarray*}
\lim_{x\to x_0+0} \psi_0(x)=+\infty
\end{eqnarray*}

(ii)  $\psi_0(x)>0$ for $x\in (x_0;+\infty)$;

(iii) for $x\in(x_0;+\infty)$ the following inequality holds:
\begin{eqnarray*}
\omega-V(x)-3\psi^2_0(x)<0
\end{eqnarray*}

Then {\em any} solution $\psi_+(x,C_+)\in S_+$ for $C_+>C^0_+$ has a
singularity at some point $x=x(C_+)>x_0$, $\lim_{x\to x(C_+)+0}
\psi_+(x,C_+)=+\infty$.
\bigskip
}

The proof of Lemma 1 is postponed in \ref{AppRepul}. Lemma 1 allows to
stop the scanning $C_+$ when $C^0_+$ satisfying the conditions of Lemma
1 is achieved. The window of scanning in this case is {\it finite}, so
one can conclude that {\it all} the solutions which exist for given
$\omega$ have been found. However, for practical implementation of
Lemma 1 in numerics one needs a condition to assure that the solution
which has been found is singular.

Let us introduce the value
\begin{eqnarray*}
\Omega\equiv\sup_{x\in{\bf R}}(\omega-V(x))
\end{eqnarray*}
and assume that $|\Omega|<\infty$. If $\Omega\leq 0$ then one can prove
that Eq.(\ref{MainEqRep}) has no localized solutions. If $\Omega> 0$
the following statement holds
\medskip

{\it \underline{Lemma 2.} Let $\Omega>0$. If a solution $\psi(x)$ of
Eq.(\ref{MainEqRep}) satisfies at some point $x=x_0$ the condition
$|\psi(x_0)|>\sqrt{\Omega}$ then this solution is singular.}
\medskip

The proof of Lemma 2 is also postponed in \ref{AppRepul}. Lemma 2
allows to stop the shooting procedure once the value
$\psi=\sqrt{\Omega}$ is achieved.

\section{Numerical results.}\label{Numeric}

\subsection{The quadratic potential.}\label{Harm}

\subsubsection{Repulsive nonlinearity}\label{HarmRep}

\begin{figure}
\includegraphics[scale=0.44]{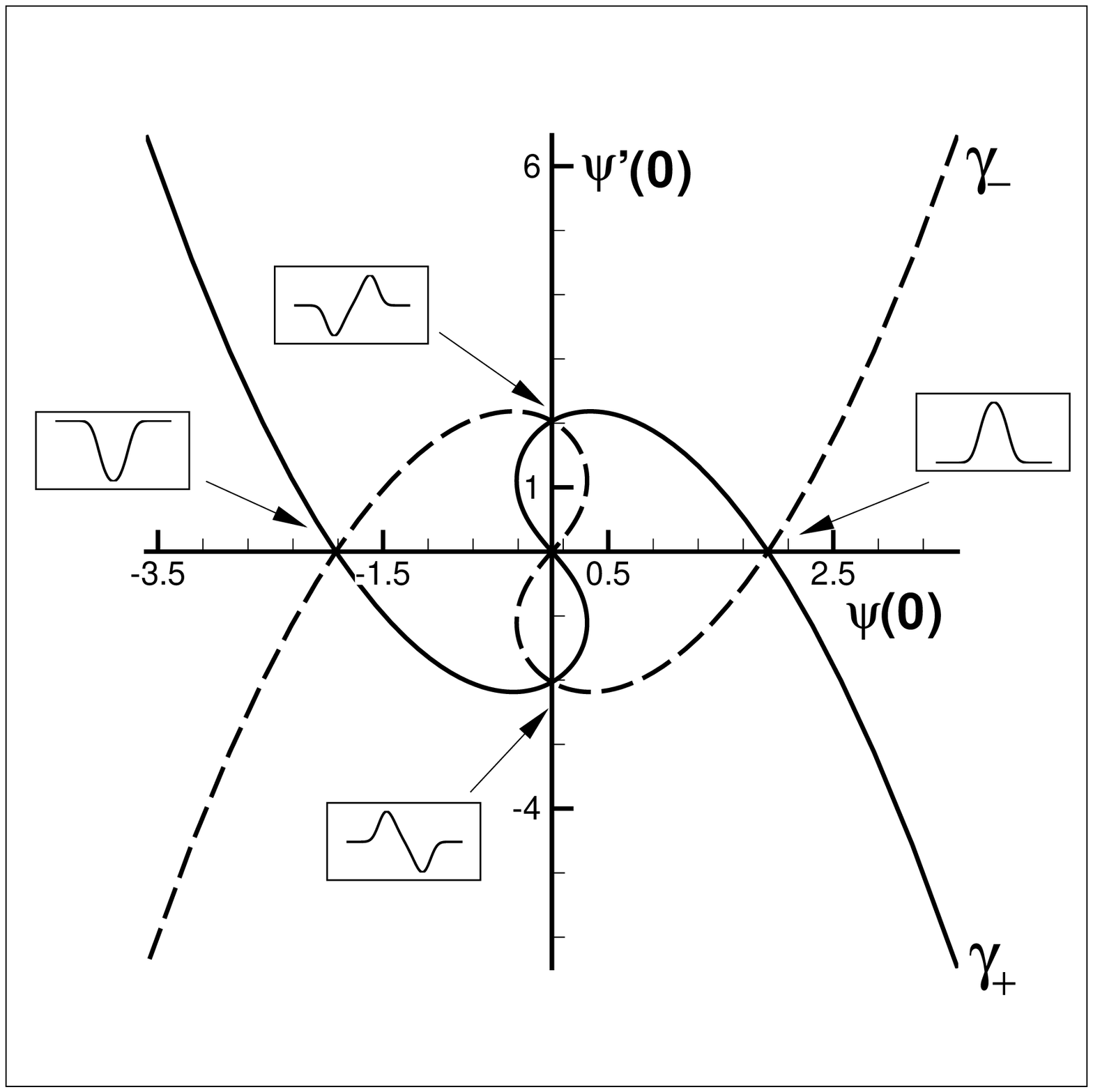}
\includegraphics[scale=0.44]{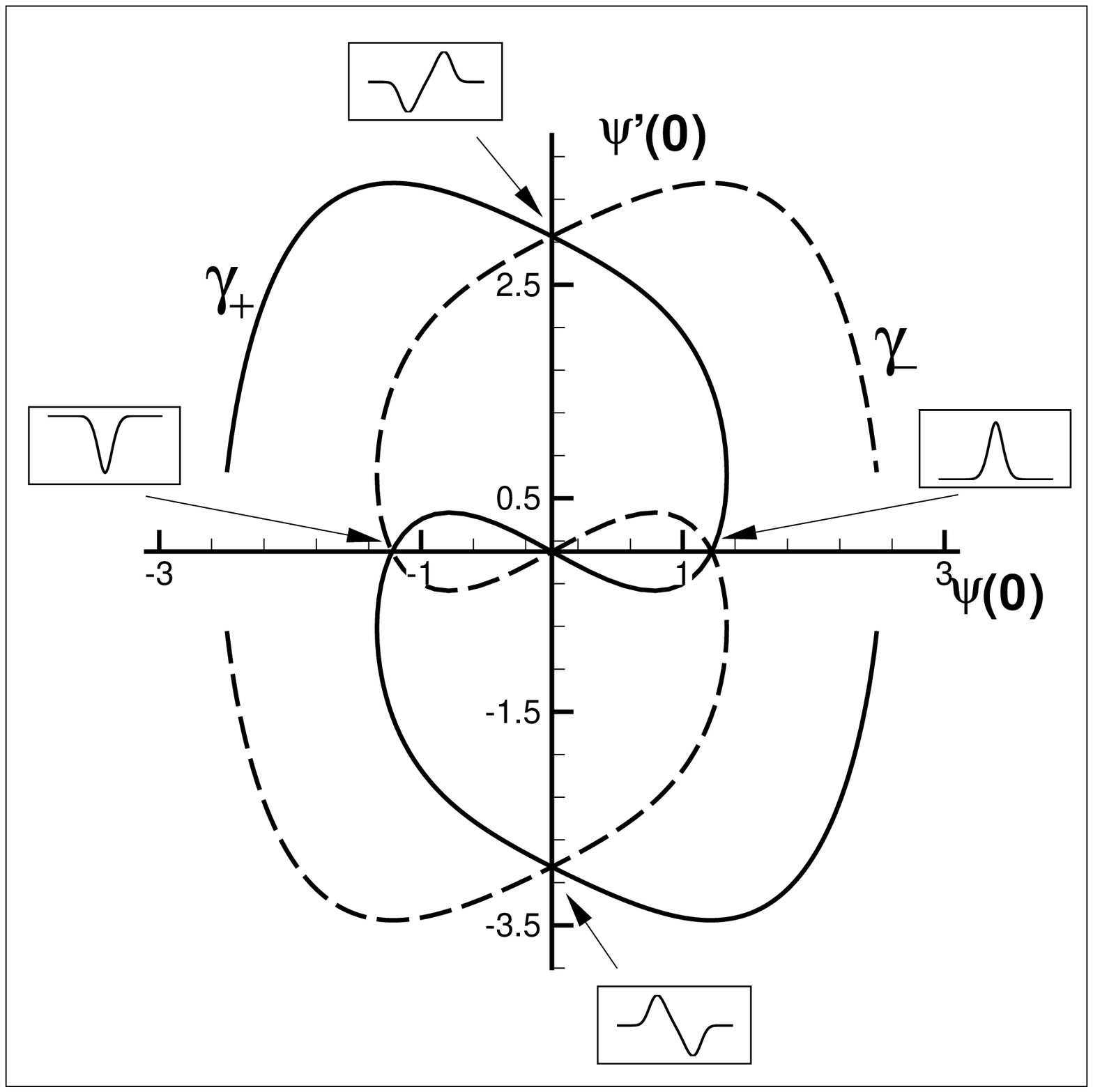}
\caption{$(\psi(0),\psi'(0))$-pictures for Eq.(\ref{MainEq}) with
$V(x)=x^2$. Left panel: repulsive nonlinearity, $\omega=4$; right
panel: attractive nonlinearity, $\omega=0$.} \label{Spirals}
\end{figure}

First, we applied the approach described above for the equation
\begin{equation}
\psi''+(\omega-x^2)\psi-\psi^3=0 \label{EqHarmRep}
\end{equation}
taking $0\leq\omega\leq 12$. According to the formula
(\ref{LinGenAsympt}) the asymptotics of the solution $\psi(x)$ as $x\to
+\infty$ is
\begin{eqnarray}
\psi(x)\sim C_+x^{\frac 12(\omega-1)}e^{-\frac{x^2}2} \label{AsHarm}
\end{eqnarray}
 The starting point for the shooting, $x_\infty$, was taken
equal to~10. The results were re-calculated for $x_\infty=12$ and no
significant change was observed. The zoom of scanning with respect to
$C_\pm$ was from $C_\pm=0$ until the value $C_\pm^{(0)}$ where the
conditions of Lemma 1 held (the value $C_\pm^{(0)}$ has been found for
all considered $\omega$). This guarantees that {\it all} localized
solutions within this range of $\omega$ were found.

\begin{figure}
\includegraphics[scale=0.44]{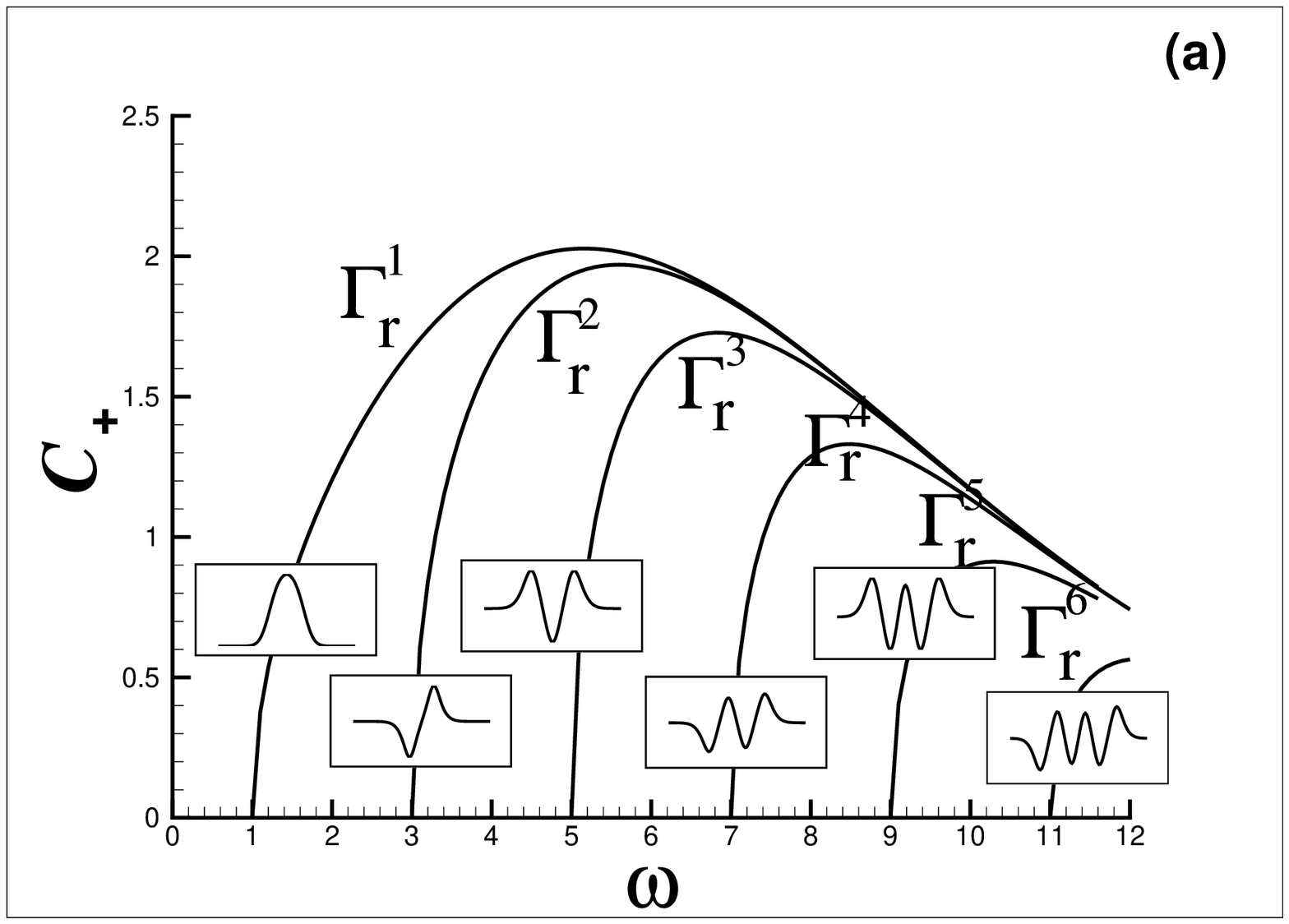}
\includegraphics[scale=0.44]{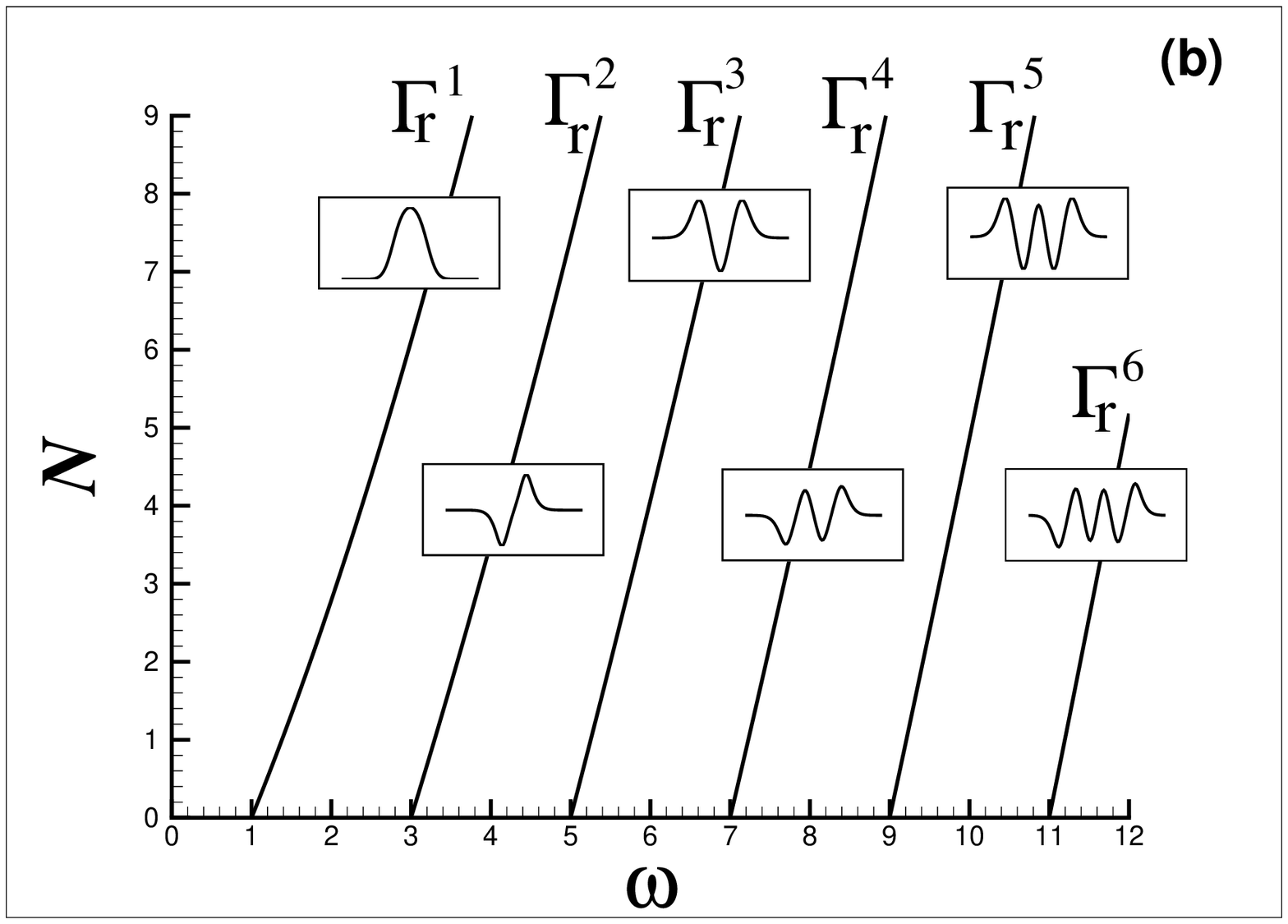}
\includegraphics[scale=0.44]{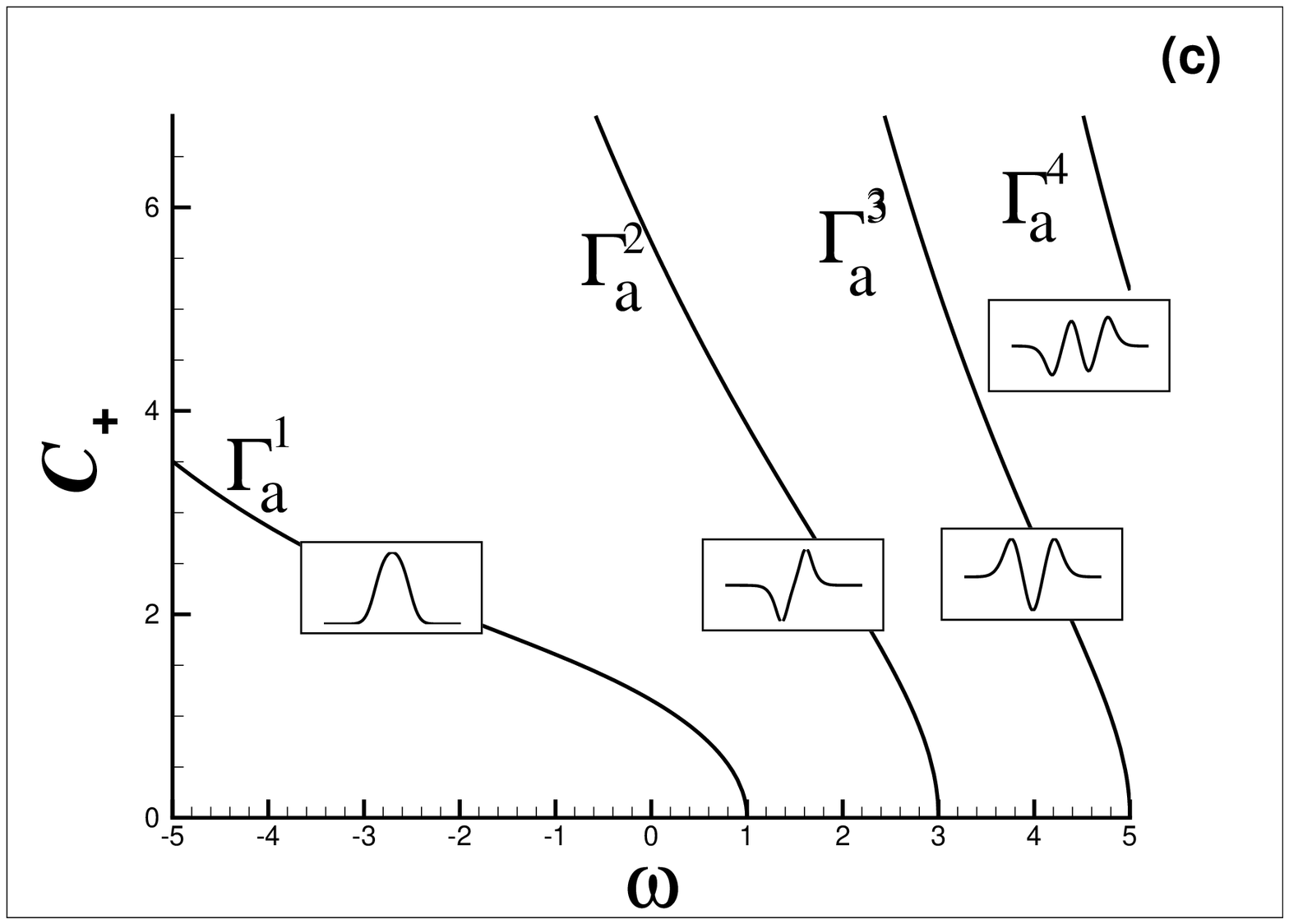}
\includegraphics[scale=0.44]{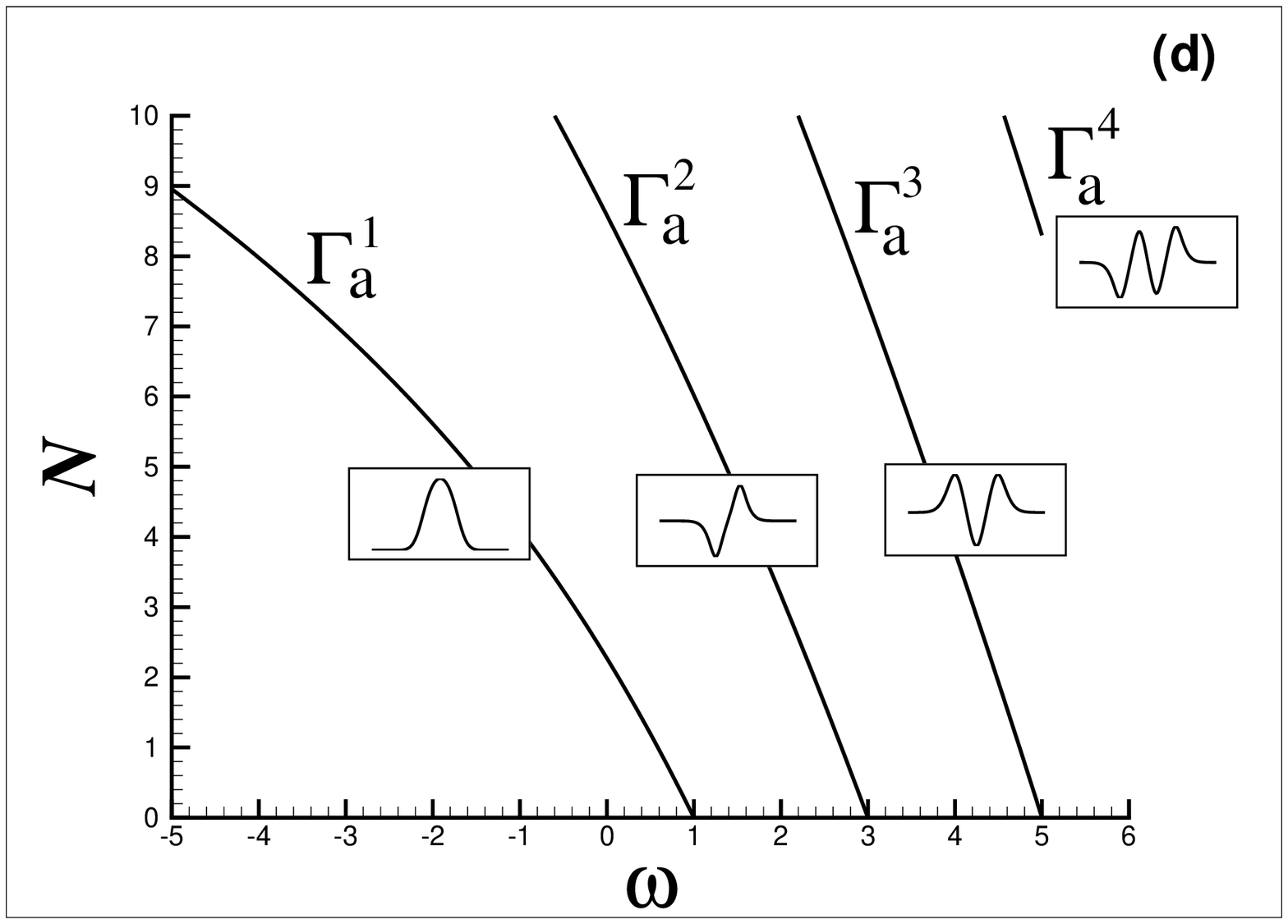}
\caption{The branches of nonlinear modes for Eq.(\ref{MainEq}) with
$V(x)=x^2$. (a): $(C_+,\omega)$-diagrams, repulsive nonlinearity; (b):
$(N,\omega)$-diagrams, repulsive nonlinearity; (c)
$(C_+,\omega)$-diagrams, attractive nonlinearity; (d)
$(N_+,\omega)$-diagrams, attractive nonlinearity. All the modes are
originated by the modes of linear harmonic oscillator.} \label{HarmFig}
\end{figure}

The Fig.\ref{Spirals}, left panel, shows the $(\psi(0),\psi'(0))$-
picture for $\omega=4$. In Fig.\ref{HarmFig}(a)-(b)  six branches of
nonlinear modes for Eq.(\ref{MainEq}) with repulsive nonlinearity are
depicted. All these branches, $\Gamma^{n}_r$, are originated by
localized modes of linear harmonic oscillator, corresponding to
eigenvalues $\omega=2n-1$, $n=1\div 6$. The linear limit can be
achieved by passing $C_+\to 0$ (or, correspondingly, $N\to 0$). The
branches of even and odd solutions alternate. It follows from these
figures that there are {\it no other branches} of localized modes
within this range of $\omega$. The analysis of linear stability for the
branches $\Gamma_{r}^n$ shows that the solutions corresponding to the
branches $\Gamma^1_{r}$ and $\Gamma^2_{r}$ are stable whereas the
solutions of the other two branches (and, probably, of all the higher
branches) are unstable. This completely agrees with results of
\cite{KevrekDyn}, (see Fig. 1 of \cite{KevrekDyn}).

\subsubsection{Attractive nonlinearity}\label{HarmAtt}
For the case of attractive nonlinearity,
\begin{equation}
\psi''+(\omega-x^2)\psi+\psi^3=0 \label{EqHarmAtt}
\end{equation}
it is proved \cite{GaussTail} that each localized mode of harmonic
oscillator, corresponding to $\omega=2n-1$, $n=1,2,\ldots$ originates a
branch of nonlinear modes, $\Gamma^n_{a}$. The asymptotics of the
solution $\psi(x)$ as $x\to +\infty$ is described by the formula
(\ref{AsHarm}). We study Eq.(\ref{EqHarmAtt}) for $-5\leq\omega\leq 5$.
Since in this case Lemma 1 cannot be applied the scanning with respect
to $C_\pm$ was fulfilled within the fixed range $0\leq C_\pm\leq 10$.
Fig.\ref{Spirals} (right panel) shows $(\psi(0),\psi'(0))$- picture for
$\omega=0$. Fig.\ref{HarmFig}(c) represents $(C_+,\omega)$-diagram and
Fig.\ref{HarmFig}(d) shows the corresponding $(N,\omega)$-diagram. The
four branches $\Gamma^n_{a}$, $n=1,2,3,4$ originated by the localized
modes of harmonic oscillator have been found and no other modes have
appeared within the considered range of $C_\pm$ and $\omega$. We
suppose that, in general, the branches of solutions generated by the
modes of linear oscillator constitute {\it all} nonlinear localized
modes that can be described by Eq.(\ref{EqHarmAtt}).

Linear stability analysis shows that the solutions of the two lowest
branches $\Gamma^1_{a}$ and $\Gamma^2_{a}$ are {\it stable} and the
solutions of the branch $\Gamma^4_{a}$ (and, probably, of all the
higher branches $\Gamma^n_{a}$, $n\geq 4$) are unstable. This conforms
with the results of \cite{KevrekDyn}, (see Fig.1 of \cite{KevrekDyn}).
At the same time we observed that the two-node solutions of the branch
$\Gamma^3_{a}$ are unstable in small-amplitude limit (for
$\omega^*<\omega<5$, $\omega^*\approx3.83$) but they become stable for
larger amplitudes ($\omega<\omega^*$). Probably this is the reason of
troubles with generation of this states by adiabatic variation of
scattering length reported in \cite{KevrekDyn}.

\subsection{The double-well potential}\label{DW}

\subsubsection{Repulsive nonlinearity.}\label{DWRep}
In this case the Eq.(\ref{MainEq}) reads
\begin{equation}
\psi''+(\omega-\alpha x^4-\beta x^2)\psi-\psi^3=0,\quad \alpha>0,\quad
\beta<0 \label{EqDWRep}
\end{equation}
According to the formula (\ref{LinGenAsympt}) the asymptotics of the
solution $\psi(x)$ as $x\to +\infty$ is
\begin{eqnarray}
\psi(x)\sim \frac{C_+}x\cdot\exp\left\{-\frac {\sqrt{\alpha}}3
x^3-\frac{\beta}{2\sqrt{\alpha}}
x-\frac{1}{2\sqrt{\alpha}}\left(\omega^2+\frac{\beta^2}{4\alpha}\right)x^{-1}\right\}
\label{AsDW}
\end{eqnarray}
All the calculations were done for the parameters of potential
$\alpha=10$, $\beta=-20$ for $-5\leq\omega\leq 10$. The range of
scanning for $C_\pm$ was from $C_\pm=0$ until the value $C_\pm^{(0)}$
where the conditions of Lemma 1 held. The value $C_\pm^{(0)}$ was found
for all $\omega$ which we considered. This implies that all nonlinear
modes within this range of $\omega$ were found.

\begin{figure}
\centerline{\includegraphics[scale=0.6]{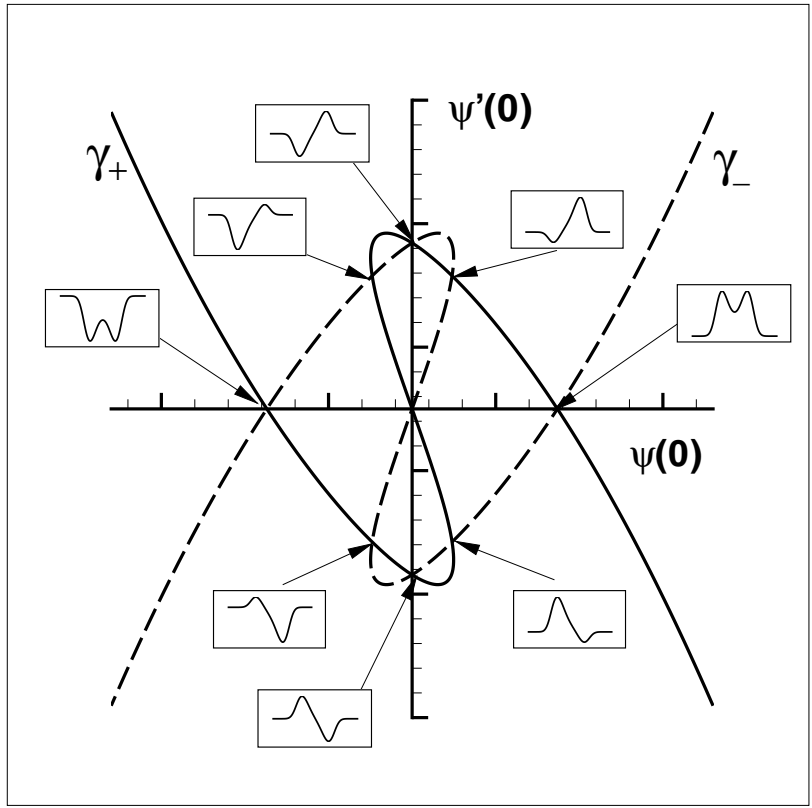}}
\caption{$(\psi(0),\psi'(0))$-pictures for Eq.(\ref{MainEq}) with
$V(x)=10x^4-20x^2$ and repulsive nonlinearity, $\omega=-3$.}
\label{SpiralsDW}
\end{figure}

\begin{figure}
\includegraphics[scale=0.44]{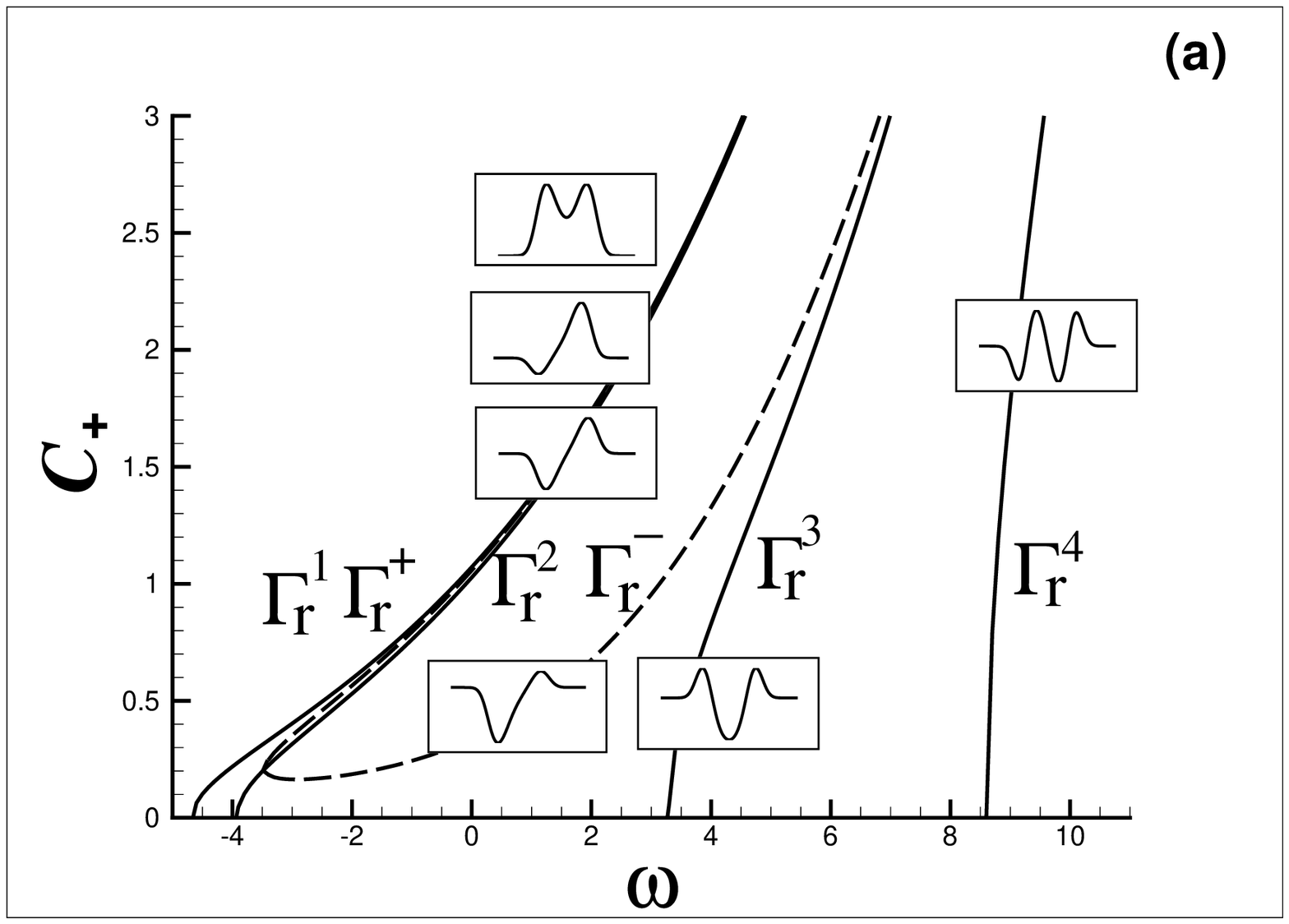}
\includegraphics[scale=0.44]{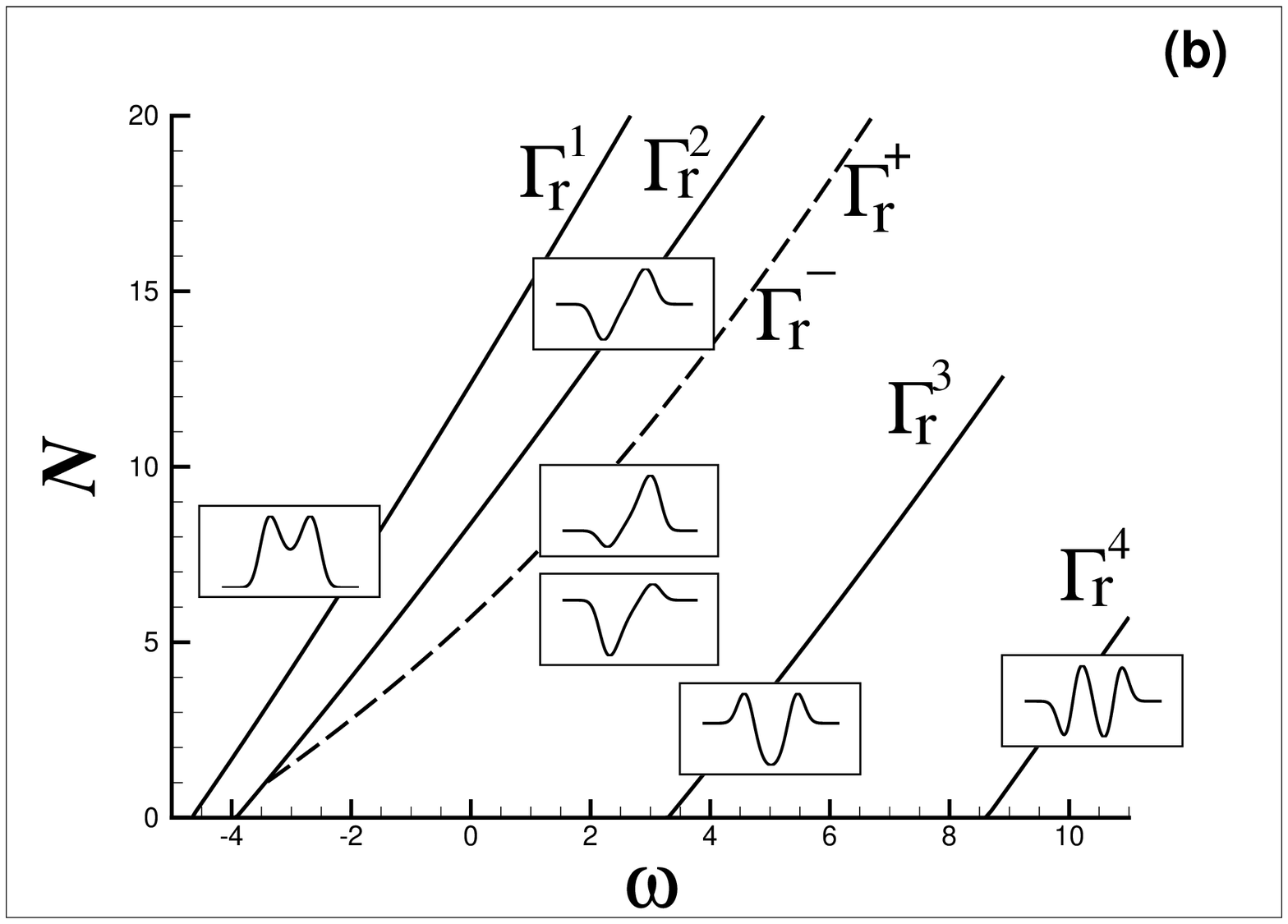}
\includegraphics[scale=0.44]{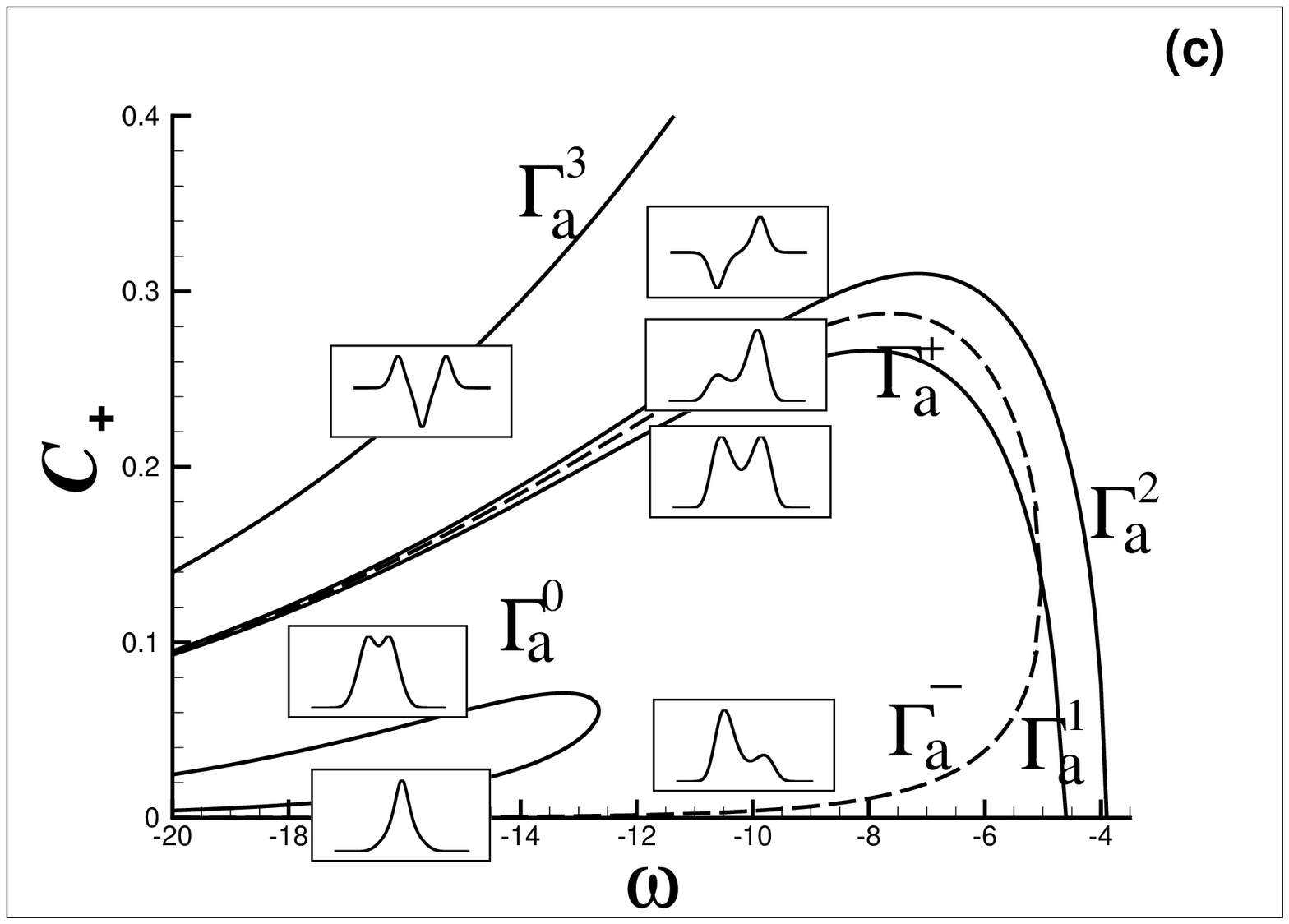}
\includegraphics[scale=0.44]{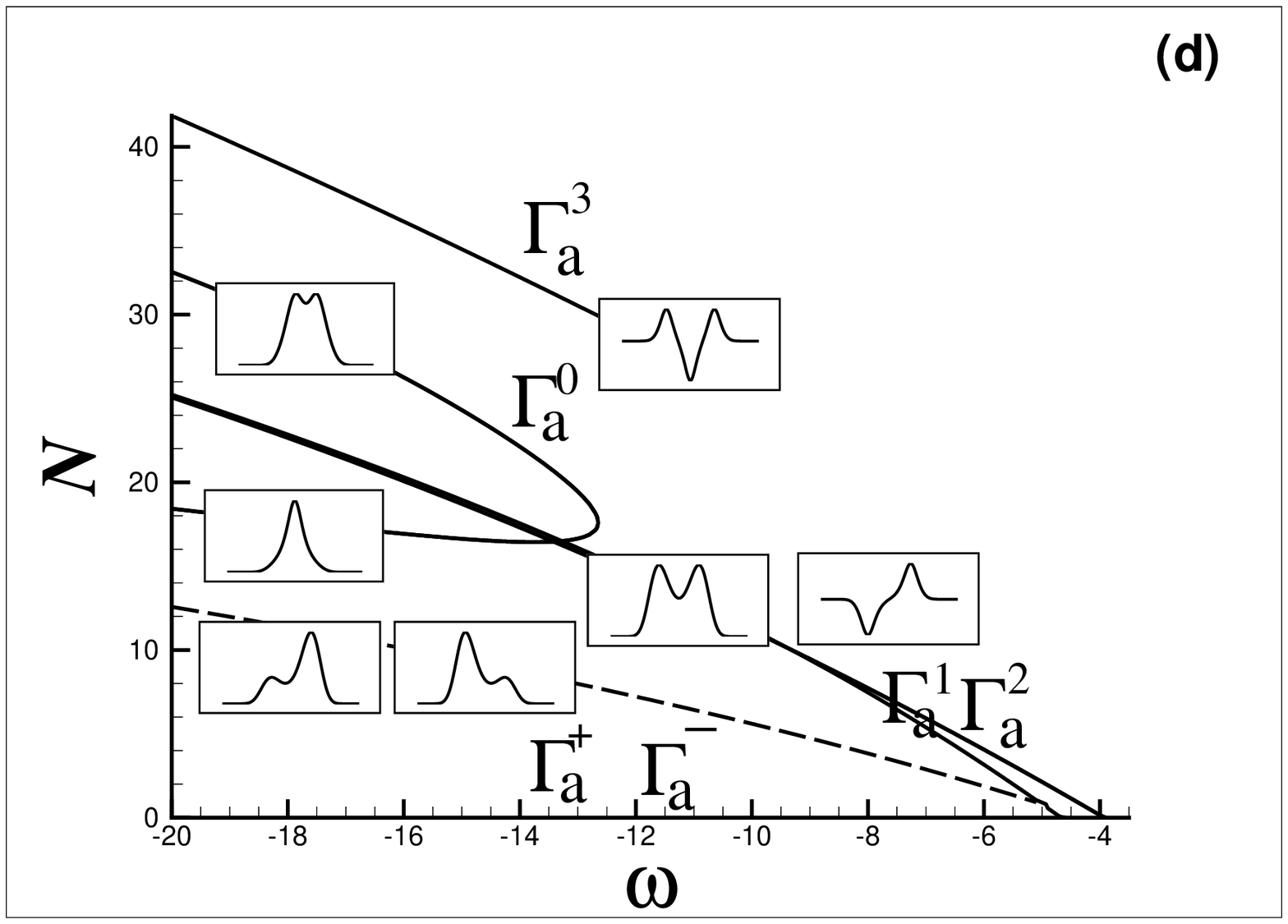}
\caption{{\it Panels (a)-(b)}: The branches of nonlinear modes for
Eq.(\ref{MainEq}) with $\alpha=10$, $\beta=-20$ and repulsive
nonlinearity. (a): $(C_+,\omega)$-diagram, (b): $(N,\omega)$-diagram.
The branches $\Gamma_r^n$, $n=1,2,3,4$ are originated by the modes of
linear eigenvalue problem, the branches $\Gamma_r^\pm$ appear as a
result of secondary branching of $\Gamma_r^2$ (the bifurcation is of
pitchfork type). In $(N,\omega)$-diagram the branches $\Gamma_r^+$ and
$\Gamma_r^-$ coincide.\\ {\it Panels (c)-(d)}: The branches of
nonlinear modes for Eq.(\ref{MainEq}) with $\alpha=10$, $\beta=-20$ and
attractive nonlinearity. (c): $(C_+,\omega)$-diagram, (d):
$(N,\omega)$-diagram. The branches $\Gamma_a^1$ and $\Gamma_a^2$ are
originated by the modes of linear eigenvalue problem, the branches
$\Gamma_a^\pm$ appear as a result of secondary branching of
$\Gamma_a^1$ (the bifurcation is of pitchfork type). In
$(N,\omega)$-diagram the branches $\Gamma_a^+$ and $\Gamma_a^-$
coincide. } \label{DWFig}
\end{figure}

Fig.\ref{SpiralsDW} represents $(\psi(0),\psi'(0))$- picture for
$\omega=-3$. Fig.\ref{DWFig}(a)-(b) shows four branches of localized
modes $\Gamma_r^n$, $n=1,2,3,4$  which depart from the eigenvalues of
linear problem (in this case no analytical form of linear modes is
available). In addition, the branch $\Gamma_r^2$ undergoes secondary
{\it pitchfork bifurcation} and two branches $\Gamma_r^\pm$ of
nonsymmetric solutions arise. These solutions are localized in one of
the wells of the potential $V(x)$.  The linear stability analysis shows
that the modes corresponding to the branches $\Gamma_r^1$,
$\Gamma_r^3$, $\Gamma_r^4$ as well as $\Gamma_r^\pm$ are stable. The
nonlinear modes corresponding to the branch $\Gamma_r^2$ are stable
until the point of pitchfork bifurcation. After the point of
bifurcation they become unstable. This is quite expectable due to the
results of the papers \cite{Wein2004,Kevrek05,KevrekPLA} and refines
the statement of \cite{AgPres2002} where it has been suggested that all
nonlinear modes in the case of repulsive nonlinearity are stable.

\subsubsection{Attractive nonlinearity.}\label{DWAtt}
In this case  Eq.(\ref{MainEq}) reads
\begin{equation}
\psi''+(\omega-\alpha x^4-\beta x^2)\psi+\psi^3=0,\quad \alpha>0,\quad
\beta<0 \label{EqDWAtt}
\end{equation}
The most part of calculations were done for $\alpha=10$, $\beta=-20$.
Fig.\ref{DWFig} represents all the branches of localized modes which
were found within the range $0\leq C_\pm\leq 0.4$ and $-20\leq
\omega\leq -4$. There are two branches which correspond to modes with
linear counterpart, $\Gamma_a^1$ and $\Gamma_a^2$ originated from 1-st
and 2-nd eigenvalues of linear problem. At some value of $\omega$ the
branch $\Gamma_a^1$ undergoes pitchfork bifurcation and two more
branches of non-symmetric solution $\Gamma_a^\pm$ emerge. Apart from
them there is a branch $\Gamma_a^0$ of nonlinear modes localized on the
maximum of the potential situated between the two wells. These
solutions were reported also in \cite{AgPres2002}. When widening the
window of scanning with respect to $C_\pm$ and $\omega$ much more
branches appear; as a result, the complete picture becomes quite
complex. The linear stability analysis shows that the solutions of the
branch $\Gamma_a^2$ as well as the branches $\Gamma_a^\pm$ are stable,
whereas the solutions of the branch $\Gamma_a^0$ are unstable. As it
was in repulsive case, the nonlinear modes corresponding to the branch
$\Gamma_a^1$ are stable until the point of pitchfork bifurcation and
are unstable after this point. The part of solutions of the branch
$\Gamma_a^3$ shown in Fig.\ref{DWFig}(c)-(d) are unstable.

\begin{figure}
\includegraphics[scale=0.42]{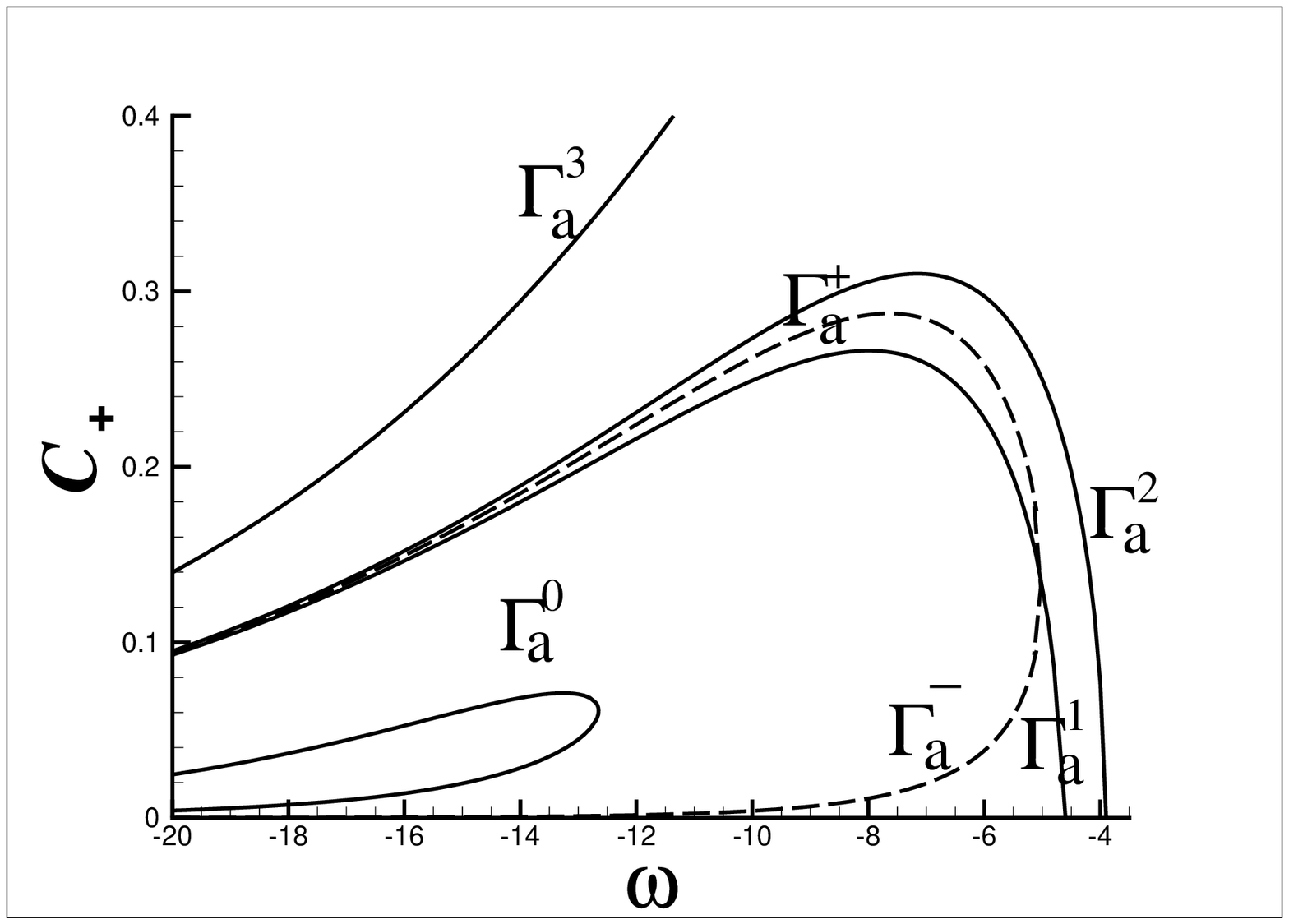}
\includegraphics[scale=0.42]{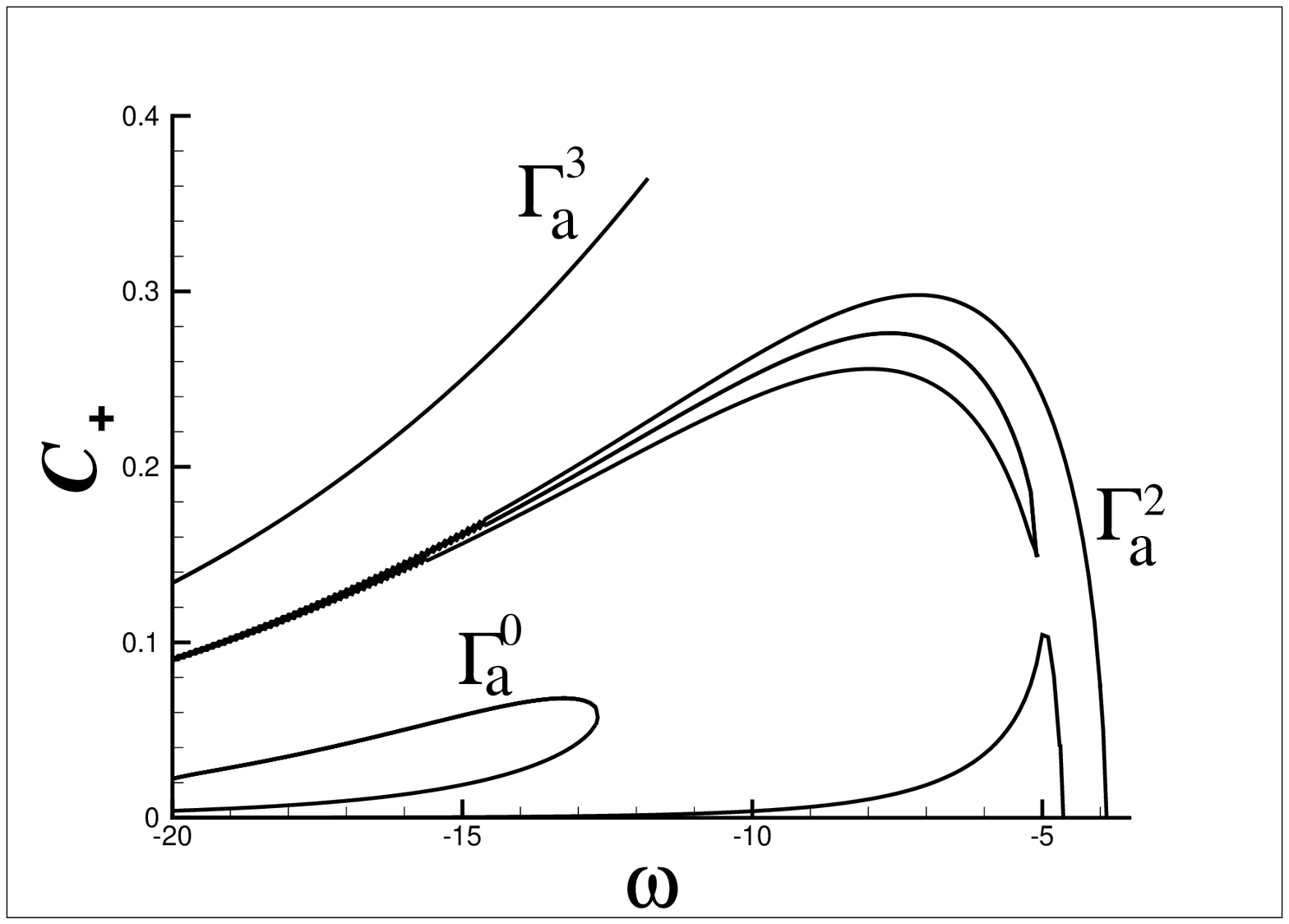}
\caption{$(C_+,\omega)$-diagrams for Eq.(\ref{MainEq}), attractive
case, $V(x)=10x^4-20x^2$ (left panel) and $V(x)=10x^4-20x^2+0.01x^3$
(right panel). Breaking of the symmetry of the potential results in
replacing of pitchfork bifurcation by saddle-node one.}
\label{Sym-Nonsym}
\end{figure}

The presence of pitchfork bifurcation reflects the fact that
Eq.(\ref{EqDWAtt}) with even potential is invariant with respect to
transformation $x\to -x$. The two nonsymmetric states $\psi^{(+)}(x)\in
\Gamma_a^+$ and $\psi^{(-)}(x)\in \Gamma_a^-$ which bifurcate from the
principal state $\psi_{1}(x)\in \Gamma_a^1$ are connected by the
symmetry $\psi^{(+)}(-x)=\psi^{(-)}(x)$. Adding small perturbation
which breaks evenness of the potential results in replacing the
pitchfork bifurcation by saddle-node one (see Fig.\ref{Sym-Nonsym}).
Similar phenomenon was reported in the paper \cite{Kevrek05} for the
case of the potential composed by harmonic and periodic parts.

\begin{figure}
\includegraphics[scale=0.28]{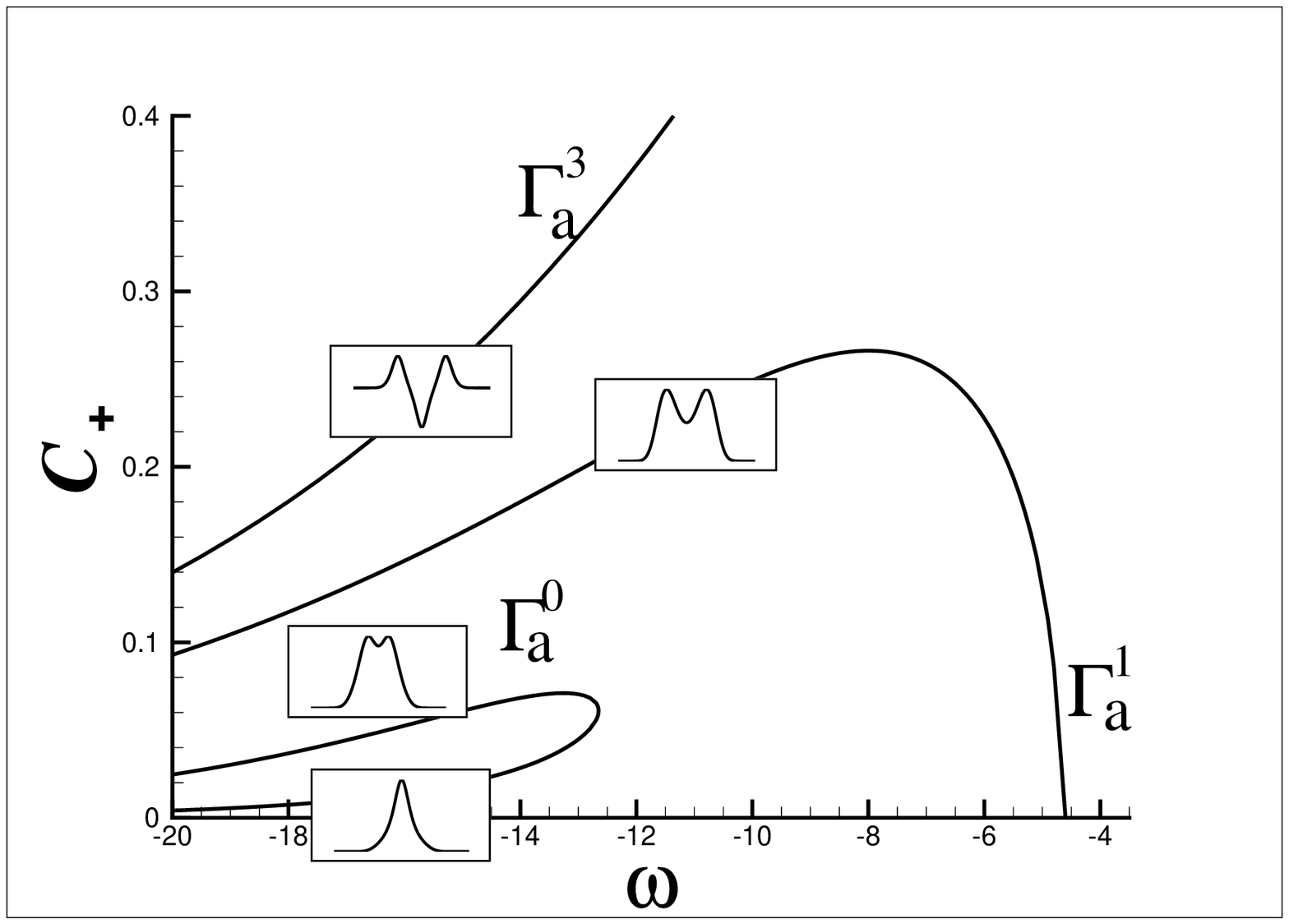}
\includegraphics[scale=0.28]{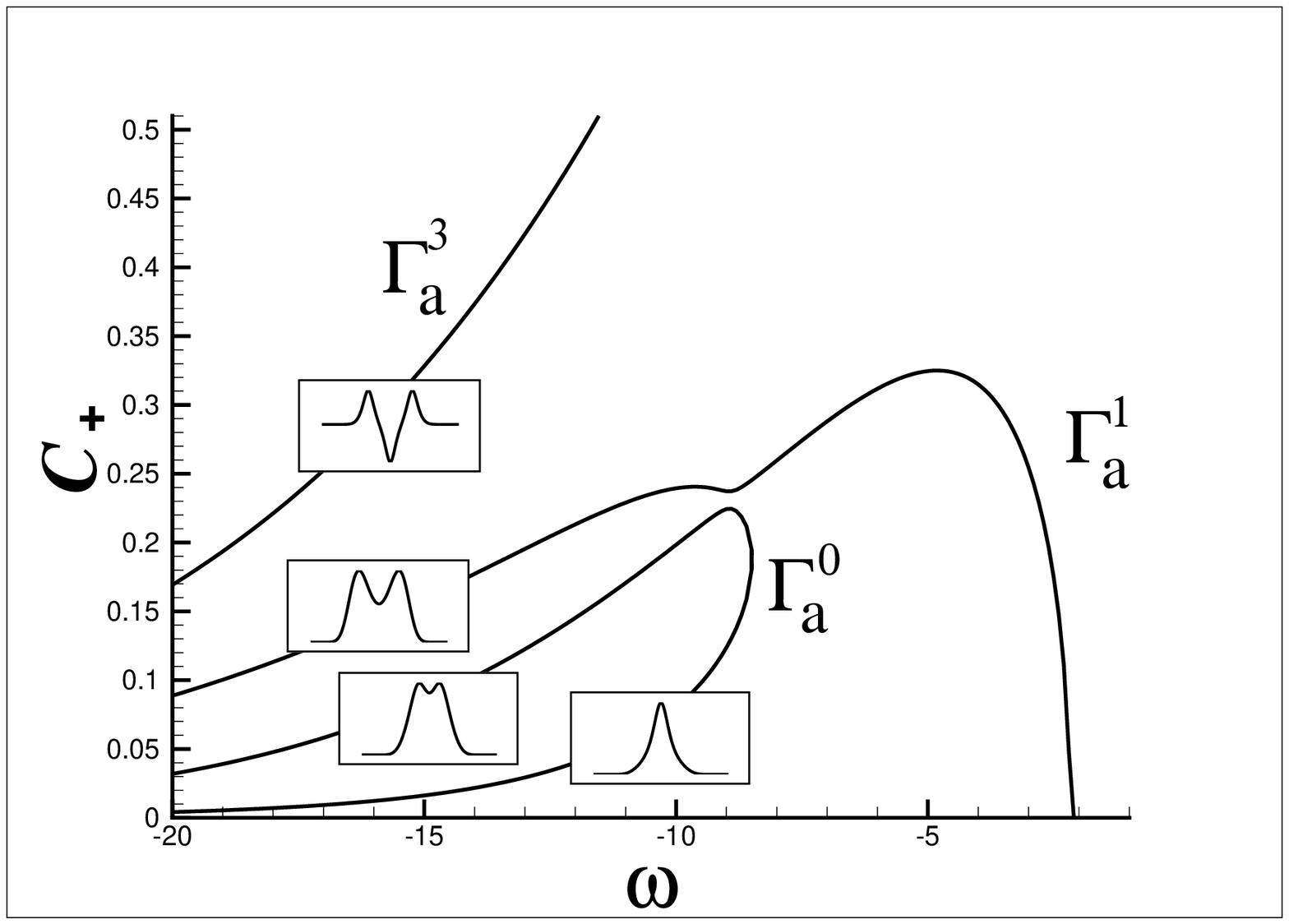}
\includegraphics[scale=0.28]{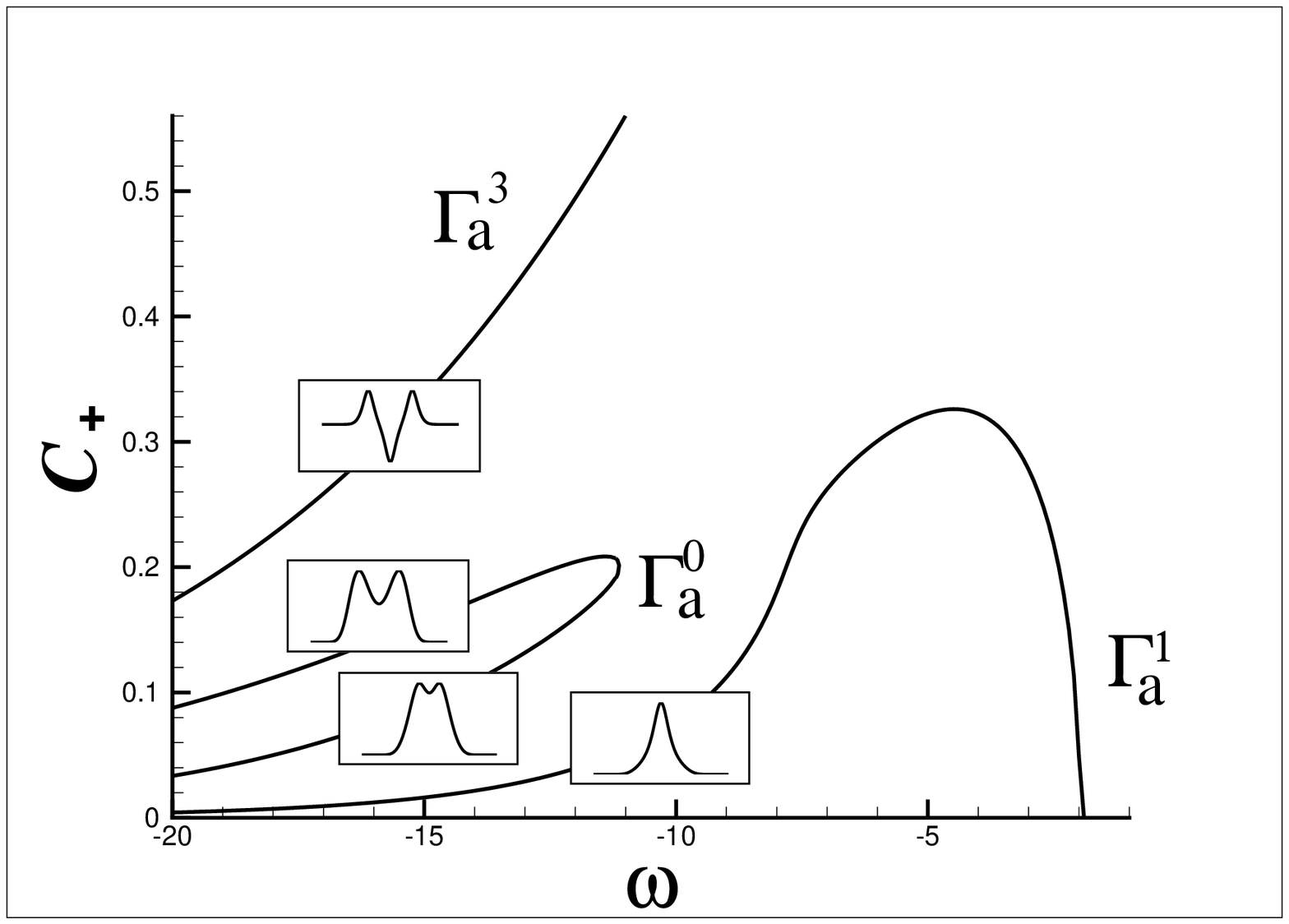}
\caption{Transformation of $(C_+,\omega)$-diagram when the parameters
of the potential vary. Left panel: $\alpha=10$, $\beta=-20$, central
panel: $\alpha=10$, $\beta=-15.91$, right panel: $\alpha=10$,
$\beta=-15.5$. Only branches of even solutions are depicted.}
\label{With-Without}
\end{figure}

One more interesting phenomenon takes place when changing the
parameters of the potential. It was found that if  $\alpha$ is fixed
there exists a critical value of $\beta$ when the branches $\Gamma_a^1$
and $\Gamma_a^0$ touch. When $\beta$ passes through this critical value
a reconnection of branches $\Gamma_a^1$ and $\Gamma_a^0$ occurs (see
Fig.\ref{With-Without}). This means that by small variation of
parameter $\beta$ (keeping $\alpha$ fixed) in vicinity of this critical
point, a solution {\it with} linear counterpart can be transformed into
a solution {\it without} linear counterpart and vice versa.

\section{Conclusion}\label{Concl}

In the paper a method to study nonlinear modes for the Gross-Pitaevskii
equation is described. It is based on exact statement on one-to-one
correspondence between the solutions of Eq.(\ref{MainEq}) which vanish
at $x\to +\infty$ and a linear coefficient $C_+$ which describe its
asymptotic in this limit. This method can be regarded as an extension
of shooting method used for calculation of single nonlinear mode (see
e.g. \cite{Rupr1995,Gammal1999}) aimed to get a {\it complete picture}
of coexisting nonlinear modes.
Combined with
some additional exact statements which have been proved in the case of
repulsive nonlinearity the method allows to fulfill demonstrative
computation to guarantee that {\it all} the nonlinear modes
corresponding to given value of $\omega$ (chemical potential in terms
of BEC) are found.

Conforming to the approach which we describe we offer a representation
of the branches of solutions on the plane $(C_+,\omega)$ instead of
more traditional  representation on the plane $(N,\omega)$,
$N=\int_{-\infty}^\infty \psi^2(x)~dx$. The advantage of this
representation is that each point of the plane $(C_+,\omega)$
corresponds to  at most one nonlinear mode whereas the point on the
plane $(N,\omega)$  can correspond to many nonlinear modes which share
the same number of particles. Therefore the intersection (or touching)
of the curves which correspond to the branches of nonlinear modes in
the plane $(C_+,\omega)$ means that some bifurcation occurs. This is
not the case for $(N,\omega)$ representation. The
$(C_+,\omega)$-diagrams calculated in the case of harmonic potential
$V(x)=x^2$ indicate that all the nonlinear modes have linear
counterpart. In the case of double-well potential the situation is
richer, due to modes without linear counterpart and secondary branching
of non-symmetric modes.

In the paper the approach is applied to 1D GPE with potential having
one or two minima. However it seems to be an interesting problem to
apply this method for the study of 1D GPE with potentials with three or
more minima (see e.g. \cite{Crowd}). Also it may be used for GPE with
periodic potential (see e.g. \cite{AKS,Kiv2003}) as well as for the
potential generated by magnetic fields and optical lattice
\cite{Kevrek05}. Applications to quintic or cubic-quintic GPE model are
also possible \cite{Fatkh,AKP}. With minor modifications this approach
can be applied to 2D and 3D radial versions of GPE.

\ack Authors are grateful to Prof. V.Konotop and Prof. L.Lerman for
fruitful discussions and pointing to useful references. G.A.
acknowledges the support from President Program in Support for Leading
Scientific Schools in Russia (Project 4122.2006.2.).

\appendix

\section{Summary of results for the equation $\psi
''-Q(x)\psi=0$.}\label{Ap1}

\newcommand{\ApA}{A}
The equation
\begin{equation}
\psi ''-Q(x)\psi=0 \label{LS}
\end{equation}
(the prime denotes the derivative with respect to $x$) has been
discussed in numerous books and papers \cite{BerShub,Fedoryuk,Kamke}.
Let us set forth below some results about Eq. (\ref{LS}) which have
been used in this paper.
\medskip

{\it \underline{Statement \ApA.1:}(see \cite{BerShub}, chapter 2) Let
there exists some $x_0$ and $\varepsilon$ such that
\begin{equation}
Q(x)\geq\varepsilon>0\quad \mbox{for}\quad x>x_0. \label{MainAssum}
\end{equation}
\medskip

Then a pair of linearly independent solutions of (\ref{LS}),
$\psi_1(x)$ and $\psi_2(x)$, can be chosen in such a way that for some
value $a>x_0$ the following assertions hold:
\medskip

(A1) the functions $\psi_1(x)$, $\psi_1'(x)\to 0$ when $x\to +\infty$;
\medskip

(A2) the functions $\psi_1(x)$, $\psi_1'(x)$ have no zeros and are
monotonic for $x>a$;
\medskip

(A3) for $x>a$
\begin{equation}
|\psi_1(x)|\leq A_1e^{-\sqrt{\varepsilon}x} \label{ExpEst2}
\end{equation}
for some $A_1$.

(A4) the solution $\psi_1(x)$ is unique (up to constant factor).
\medskip

(B1) the functions $\psi_2(x),\psi_2'(x)\to \infty$ when $x\to
+\infty$;
\medskip

(B2) the functions $\psi_2(x),\psi_2'(x)$ have no zeros and are
monotonic for $x>a$;
\medskip

(B3) for $x>a$
\begin{equation} |\psi_2(x)|\geq
A_2e^{\sqrt{\varepsilon}x} \label{ExpEst1}
\end{equation}
for some $A_2$.
\medskip
}

\underline{\it Comment:} The points (A1)-(A3) and the fact that the
sign of $\psi''(x)>0$ coincides with the sign of $\psi(x)$ for
$[a;+\infty)$ imply that
\begin{eqnarray}
\psi_1(x),\psi_1'(x),\psi_1''(x)\in L_p[a;+\infty);\quad p>0
\label{Summ1}
\end{eqnarray}

 Imposing more restrictions to $Q(x)$ results in more
precise description for the solutions $\psi_1(x)$ and $\psi_2(x)$
growth or decay:
\medskip

{\it \underline{Statement \ApA.2:}(see \cite{Coppel}, Chapter IV,
Theorem 14) Let the function $Q(x)$ satisfies the condition
(\ref{MainAssum}). Let, in addition, the following conditions hold:
\medskip

(a) $Q(x)\in C^2(x_0,+\infty)$;
\medskip

(b) $\int_{x_0}^\infty \frac{|Q''(x)|}{Q^{3/2}(x)}~dx<\infty,$
\medskip

Then the solutions $\psi_1(x)$, $\psi_2(x)$  (which exist due to
Statement \ApA.1) obey the following asymptotic formulas:

\begin{eqnarray}
&&\psi_1(x)= \frac {e^{-S(x_0,x)}}{Q^{1/4}(x)}\left(1
+o(1)\right),\label{Asymptot-}\\
&&\psi_2(x)= \frac {e^{S(x_0,x)}}{Q^{1/4}(x)}\left(1 +o(1)\right),
\label{Asymptot+} \quad S(x_0,x)= \int\limits_{x_0}^x\sqrt{Q(t)}dt
\end{eqnarray}
Moreover, the asymptotic formulas (\ref{Asymptot+})-(\ref{Asymptot-})
can be differentiated at least once.}

\underline{\it Comment:} Due to (\ref{MainAssum}) $S(x_0,x)\to+\infty$
as $x\to +\infty$.
\medskip

\underline{\it Comment:} Evidently, there is a freedom in choosing the
lower bound of integration  in (\ref{Asymptot+})-(\ref{Asymptot-}). Its
change from $x=x_0$  to $x=x_1$ results in rescaling of $\psi_1(x)$ and
$\psi_2(x)$
\begin{eqnarray*}
\psi_1(x)\longrightarrow e^{S(x_0,x_1)}\psi_1(x); \quad
\psi_2(x)\longrightarrow e^{-S(x_0,x_1)}\psi_2(x)
\end{eqnarray*}

\section{Proof of Theorem from Section \ref{back_basis}.}\label{Ap2}
\newcommand{\ApB}B

The aim of this Appendix is to prove the following statement for the
equation
\begin{equation}
\psi''-Q(x)\psi+F(\psi)=0 \label{NonlS}
\end{equation}

{\it \underline{Theorem 1.} Let there exists some $x_0$ such that
\medskip

(Q1) $Q(x)\geq\varepsilon>0$ for $x>x_0$;
\medskip

(Q2) $Q(x)\in C^2(x_0,+\infty)$;
\medskip

(Q3) $\int_{x_0}^\infty \frac{|Q''(x)|}{Q^{3/2}(x)}~dx<\infty$
\medskip

and there exist some $\delta$ such that
\medskip

(NL1) $F(t)\in C^3[-\delta;\delta]$.
\medskip

(NL2) $F(0)=F'(0)=0$;
\medskip

Let $\psi_1(x)$ be the solution of linear equation $\psi''-Q(x)\psi=0$
such that $\psi_1(x)\to 0$ as $x\to+\infty$ (see \ref{Ap1}). Let $S_+$
be a set of all solutions of Eq.(\ref{NonlS}) such that $\psi(x)\to 0$
when $x\to+\infty$. Then
\medskip

(a) for any solution $\psi(x)\in S_+$ there exists a limit
\begin{equation}
\lim_{x\to +\infty}\frac{\psi(x)}{\psi_1(x)}=C<\infty
\label{Ass11}
\end{equation}

(b) for any $C\in{\bf R}$  there exists an unique solution
$\psi_+(x;C)\in S_+$ of Eq.(\ref{NonlS}) such that
\begin{equation}
\lim_{x\to +\infty}\frac{\psi_+(x,C)}{\psi_1(x)}=C \label{Ass22}
\end{equation}

(c) if  $\psi_+(x;C_+)$ exists in some neighbourhood $U$ of a
point $(\tilde x;\tilde{C}_+)$ then in $U$ there also exists
partial derivative $\frac{\partial\psi_+(x;C_+)}{\partial C_+}$
which is continuous with respect to $C_+$.}
\bigskip

\underline{\it Comment}: If for a function $F(t)$ the condition (NL2)
holds and $F(t)\in C^2[-\delta;\delta]$ for some $\delta>0$ then
$f(t)\equiv F(t)/t$ has no singularities in $|t|<\delta$ and there
exists $A_0>0$ and $\delta_1$ such that
\begin{equation}
|f(t)|\leq A_0 |t|,\quad |t|<\delta_1; \label{Useful}
\end{equation}
\bigskip

The proof of Theorem 1 can be sketched in a form of a sequence of
Lemmas. First Lemma states that small perturbations of potential in
 Eq.(\ref{LS}) do not change the asymptotic (\ref{Asymptot-})
of the solution  which tends to zero as $x\to +\infty$.
\medskip

{\it \underline {Lemma \ApB.1} Let $Q(x)$ satisfy the conditions
(Q1)-(Q3) of Theorem 1 for $x>x_0$ and $\psi_1(x)$ be a solution
of equation $\psi''-Q(x)\psi=0$ such that $\psi_1(x)\to 0$ as
$x\to +\infty$ (guaranteed by Statement \ApA.1). Let $R(x)\in
C^2(x_0,+\infty)$, $R(x)\to 0$ when $x\to +\infty$ and
$R(x),R''(x)\in L_1(x_0,+\infty)$. Then the equation
\begin{eqnarray}
\eta''-(Q(x)+R(x))\eta=0 \label{LinPerturbed}
\end{eqnarray}
has unique (up to constant factor) solution $\eta_1(x)$ such that:
\medskip

(i) $\eta_1(x)$, $\eta_1'(x)$  are monotonous functions on some
interval $[a;+\infty)$, $a\geq x_0$;
\medskip

(ii) $\eta_1(x)$, $\eta_1'(x)$ tend to zero when $x\to+\infty$;
\medskip

(iii) the solution $\eta_1(x)$ when $x\to +\infty$ has the asymptotic
\begin{equation}
\eta_1(x)=\psi_1(x)(C_1+o(1));\quad \eta'_1(x)=\psi'_1(x)(C_1+o(1))
\label{Asymptot1}
\end{equation}
for some $C_1\in {\bf R}$.
\medskip }

\underline{\it Proof:}  It is straightforward to check that
$Q_1(x)\equiv Q(x)+R(x)$ also satisfies the conditions of Statement
\ApA.2. Then the statement of Lemma \ApB.1 follows immediately from
formula (\ref{Asymptot-}). $\blacksquare$
\medskip

{\it \underline {Lemma \ApB.2} Let $Q(x)$ satisfies the conditions
(Q1)-(Q3) of Theorem 1. Let  $F(t)\in C^2[-\delta,\delta]$ for some
$\delta>0$ and satisfies  the condition (NL2). Then for any solution
$\psi(x)$ of Eq.(\ref{NonlS}) such that $\psi(x)\to 0$ as $x\to+\infty$
there exists some $C\in {\bf R}$ such that the asymptotic of $\psi(x)$
is given by formula
\begin{equation}
\psi(x)= \frac {e^{-S(x_0,x)}}{Q^{1/4}(x)}\left(C +o(1)\right),\quad
S(x_0,x)= \int\limits_{x_0}^x\sqrt{Q(t)}dt\quad
x\to+\infty\label{MainAss}
\end{equation}
The case $C=0$ corresponds to zero solution. This asymptotic formula
can be differentiated at least once. }
\bigskip

\underline{\it Proof:} Due to the condition (NL2) one can introduce
$f(t)=F(t)/t$ such that $f(0)=0$. Suppose that $\psi(x)$ is a solution
of Eq.(\ref{NonlS}) such that $\psi(x)\to 0$ as $x\to+\infty$. Let us
rewrite Eq.(\ref{NonlS}) in the form
\begin{eqnarray*}
\psi''-(Q(x)+R(x))\psi=0,\quad R(x)=-f(\psi(x));
\end{eqnarray*}
One can show that the conditions of Lemma \ApB.2 imply the conditions
of Lemma \ApB.1 for  $R(x)$. Then the statement of Lemma \ApB.2 follows
from Lemma \ApB.1. $\blacksquare$
\medskip

{\it \underline {Lemma \ApB.3} Let $Q(x)$ satisfies the conditions
(Q1)-(Q3) for some $x>x_0$. Let $F(t)\in C[-\delta,\delta]$ for
some $\delta>0$ and $F(t)$ satisfies the condition (\ref{Useful})
for $|t|<\delta$. Let $\psi_1(x)$ be some fixed nonzero solution
of linear problem (\ref{LS}), such that $\psi_1(x)\to 0$ when
$x\to+\infty$. Then for any $C\in {\bf R}$ there exists a solution
$\psi_+(x;C)$ of (\ref{NonlS}) such that
\begin{equation}
\lim_{x\to+\infty}\frac{\psi_+(x;C)}{\psi_1(x)}=C \label{AnyC}
\end{equation}
}
\bigskip

 \underline{\it Comment}. This statement is converse to Lemma \ApB.2.
The assertion of Lemma \ApB.3 says that for any $C$ there exists a
solution of (\ref{NonlS}) which at $x\to+\infty$ has the asymptotic
\begin{equation}
\psi_+(x)=\frac {e^{-S(x_0,x)}}{Q^{1/4}(x)}\left(C+o(1)\right),\quad
S(x_0,x)= \int\limits_{x_0}^x\sqrt{Q(t)}dt \label{Asymptot6}
\end{equation}
\bigskip

{\it \underline{Proof:}} Let $\psi(x)$ satisfy the equation
(\ref{NonlS}). Introduce the function
\begin{eqnarray*}
\Psi(x)=\frac{\psi(x)}{\psi_1(x)}-C
\end{eqnarray*}
where $C$ is free real parameter and $\psi_1(x)$ is fixed solution of
linear equation (\ref{LS}), such that $\psi_1(x)\to 0$ as
$x\to+\infty$. Then $\Psi(x)$ satisfies the system of equations
\begin{eqnarray}
&&\Psi'=\Phi\label{Sys1}\\
&&\Phi'=-P(x)\Phi-(\Psi+C)\cdot f((\Psi+C)\psi_1(x)),\qquad P(x)=\frac
{2\psi_1'(x)}{\psi_1(x)} \label{Sys2}
\end{eqnarray}
where $f(t)=F(t)/t$. The existence of solution $\Psi(x)$ such that
$\Psi(x)\to 0$ as $x\to+\infty$ follows from Theorem 9.1, Chapter XII
of \cite{Hartman} applied to the system (\ref{Sys1})-(\ref{Sys2}). The
conditions of this Theorem can be verified straightforwardly
$\blacksquare$.

\bigskip

The following Lemma says that for given $C$ the solution of
(\ref{NonlS}) with constant $C$ in asymptotic (\ref{MainAss}) is
unique.
\bigskip

{\it \underline {Lemma \ApB.4} Let $Q(x)$ satisfies the conditions
(Q1)-(Q3) of Theorem 1 for $x>x_0$ and $F(t)$ satisfies the conditions
(NL1)-(NL2). Then for any two solutions, $\psi_a(x)$ and $\psi_b(x)$,
of Eq.(\ref{NonlS}) such that $\psi_a(x),\psi_b(x)\to 0$ when
$x\to\infty$ and $\psi_a(x)\not\equiv\psi_b(x)$ the following
inequality holds:
\begin{equation}
\lim_{x\to +\infty}\frac{\psi_a(x)}{\psi_b(x)}\ne 1 \label{Asymptot3}
\end{equation}
\medskip
}

\underline{\it Proof:} Assume the contrary, that $\lim_{x\to
+\infty}\frac{\psi_a(x)}{\psi_b(x)}=1$. Then, according to Lemma
\ApB.2, both solutions $\psi_a(x)$ and $\psi_b(x)$ share the same
asymptotics
\begin{eqnarray}
\psi_a(x)=\frac {e^{-S(x_0,x)}}{Q^{1/4}(x)}\left(C+o(1)\right),\qquad
\psi_b(x)=\frac {e^{-S(x_0,x)}}{Q^{1/4}(x)}\left(C+o(1)\right),\label{Asymptot4}\\
S(x_0,x)= \int\limits_{x_0}^x\sqrt{Q(t)}dt
\end{eqnarray}
Subtracting the second equation from the first one in the system
\begin{eqnarray*}
\psi_b''-Q(x)\psi_b+F(\psi_b)=0,\quad \psi_a''-Q(x)\psi_a+F(\psi_a)=0
\end{eqnarray*}
we conclude that the difference $\Delta=\psi_b-\psi_a$ obeys the
equation
\begin{eqnarray}
\Delta''-\left(Q(x)+G(\psi_a(x),\psi_b(x))\right)\Delta=0,\label{AuxEq1}\\[1mm]
G(\psi_a(x),\psi_b(x))=-\frac{F(\psi_b(x))-F(\psi_a(x))}{\psi_b(x)-\psi_a(x)}
\end{eqnarray}
Let us denote $R(x)=G(\psi_a(x),\psi_b(x))$. It is tedious but
straightforward to show that $R(x)$ satisfies the conditions of Lemma
\ApB.1 (the proof involves Lagrange theorem). Then  it follows from
Lemma \ApB.1 that
\begin{equation}
\Delta(x)=\frac {e^{-S(x_1,x)}}{Q^{1/4}(x)}\left(C_1+o(1)\right),\quad
S(x_1,x)= \int\limits_{x_1}^x\sqrt{Q(t)}dt \label{Asymptot5}
\end{equation}
Since $\Delta\ne 0$ then $C_1\ne 0$ and
$\lim_{x\to+\infty}\frac{\Delta(x)}{\psi_b(x)}=\frac{C_1}C\ne 0$ that
proves Lemma \ApB.4. $\blacksquare$

\bigskip

Combining the results of Lemmas \ApB.2, \ApB.3, \ApB.4 we arrive at the
statements (a) and (b) of Theorem 1. The statement (c) follows from the
uniqueness of the solution $\psi_+(x;C)$ for given $C$ (the statement
(b)) and the property (NL1)(see e.g. \cite{Hartman}). This completes
the proof of Theorem 1.  $\blacksquare$

\section{Proof of Lemmas 1 and 2 from Section
\ref{rep_non_stat}}\label{AppRepul}

{\it \underline{Comparison Lemma.} Let the functions $y(t)$ and $x(t)$,
$t\in[a;b]$ be solutions of equations
\begin{eqnarray}
&&y_{tt}-g(t,y)=0\label{LemEq1}\\
&&x_{tt}-f(t,x)=0\label{LemEq2}
\end{eqnarray}
correspondingly. Let also the following conditions hold:
\medskip

(i) $f(t,\xi)$, $g(t,\xi)$ are defined on $[a;b]\times [A;B]$ and
locally Lipschitz continuous (see \cite{Herz2}) with respect to $\xi$,
$\xi\in [A;B]$, ($A$,$B$, $b$ may be finite or infinite);
\medskip

(ii) $g(t,\xi)\geq f(t,\xi)$ for any $t\in [a;b]$, $\xi\in [A;B]$;
\medskip

(iii) $f(t,\xi)$ is monotone nondecreasing with respect to $\xi$,
$\xi\in [A;B]$.
\medskip

Let $A<x(a)\leq y(a)<B$ and $x_t(a)\leq y_t(a)$. Then $x_t(t)\leq
y_t(t)$ and $x(t)\leq y(t)$ while $A<x(t),y(t)<B$ or for the whole
interval $t\in[a;b]$}
\medskip

{\it Proof:} Introduce the functions
\begin{eqnarray*}
\tilde{f}(t,\xi)=\left\{
\begin{array}{l}
    f(t,A),\quad \xi< A \\
    f(t,\xi),\quad A\leq\xi\leq B\\
    f(t,B),\quad \xi>B
\end{array}
\right. \qquad%
\tilde{g}(t,\xi)=\left\{
\begin{array}{l}
    g(t,A),\quad \xi< A \\
    g(t,\xi),\quad A\leq\xi\leq B\\
    g(t,B),\quad \xi>B
\end{array}
\right.
\end{eqnarray*}
and let $\tilde{x}(t)$,$\tilde{y}(t)$ be the solutions of Cauchy
problem for the equations
\begin{eqnarray*}
&&\tilde{y}_{tt}-\tilde{g}(t,\tilde{y})=0\\
&&\tilde{x}_{tt}-\tilde{f}(t,\tilde{x})=0
\end{eqnarray*}
such that $\tilde{x}(a)=x(a)$, $\tilde{x}_t(a)=x_t(a)$,
$\tilde{y}(a)=y(a)$, $\tilde{y}_t(a)=y_t(a)$. The functions
$\tilde{x}(t)$ and $\tilde{y}(t)$ coincide with $x(t)$, $y(t)$ when
$A<x(t),y(t)<B$. Consider vectors ${\bf v}, {\bf u}$ and
vector-functions ${\bf G}(t;{\bf v})$, ${\bf F}(t,{\bf u})$ defined as
\begin{eqnarray*}
&&{\bf v}=\left(
\begin{array}{c}
\tilde{y}\\ \tilde{y}_t
\end{array}
\right);\quad {\bf G}(t,{\bf v})=\left(
\begin{array}{c}
\tilde{y}_t\\ \tilde{g}(t,\tilde{y})
\end{array}
\right);\\
&&{\bf u}=\left(
\begin{array}{c}
\tilde{x}\\ \tilde{x}_t
\end{array}
\right); \quad {\bf F}(t,{\bf u})=\left(
\begin{array}{c}
\tilde{x}_t\\ \tilde{f}(t,\tilde{x})
\end{array}
\right)
\end{eqnarray*}
For ${\bf u_1},{\bf u_2}\in {\bf R}^2$ we define the order
relation: ${\bf u_1}\leq {\bf u_2}$ if each component of ${\bf
u_2}$ is greater or equal than the corresponding component of
${\bf u_1}$. The function ${\bf F}(t,{\bf u})$ is monotone
nondecreasing in the sense that if ${\bf u_1}\leq {\bf u_2}$ then
${\bf F}(t,{\bf u_1})\leq{\bf F}(t,{\bf u_2})$. Then
\begin{eqnarray*}
{\bf 0}={\bf u}_t-{\bf F}(t,{\bf u})={\bf v}_t-{\bf G}(t,{\bf v})\leq
{\bf v}_t-{\bf F}(t,{\bf v})
\end{eqnarray*}
In this situation Theorem 2 of \cite{Herz1} (see also \cite{Herz2})
states that ${\bf u}(a)\leq {\bf v}(a)$ implies that ${\bf u}(t)\leq
{\bf v}(t)$ for $t\in[a;b]$. This gives the assertion of Comparison
Lemma. $\blacksquare$
\medskip

{\it Proof of Lemma 1}. Let us denote
$u(x,C)=\frac{\partial\psi_+(x;C)}{\partial C}$. Then $u(x,C)$
satisfies the linear equation
\begin{equation}
u''+(\omega-V(x)-3\psi_+^2(x,C))u=0\label{Varia}
\end{equation}
The conditions of Lemma 1 imply that there is an interval
$[C_+^0,C_+^1)$ such that for any $C\in [C_+^0,C_+^1)$ and any $x$ for
which the solution $\psi_+(x,C)$ is defined, the following conditions
hold
\begin{eqnarray}
&&\psi_+(x;C)>0\label{Lemma1_1}\\
&&\omega-V(x)-3\psi_+^2(x;C)<0\label{Lemma1_2}
\end{eqnarray}
Let at least one of the conditions (\ref{Lemma1_1})-(\ref{Lemma1_2})
breaks at $C=C_+^1$ i.e. one at least one of the signs ``$<$'', ``$>$''
in (\ref{Lemma1_1})-(\ref{Lemma1_2}) becomes ``$=$'' at some value
$x=a$. It follows from asymptotics (\ref{GenAsympt+}) that there exists
$b$ large enough that $\psi_+(x;C)$ for $x>b$ and $C\in[C_+^0,C_+^1]$
is monotonic with respect to $C$ in the sense that if $C_1\geq C_2$,
$C_{1,2}\in[C_+^0,C_+^1]$ and $x>b$ then
$\psi_+(x;C_1)\geq\psi_+(x;C_2)$ and
$\psi_+'(x;C_1)\leq\psi_+'(x;C_2)$. This means that for $x>b$ and
$C\in[C_+^0,C_+^1]$ one has $u(x;C)\geq 0$ and $u'(x;C)\leq 0$. After
changing $x\to -x$ one can apply Comparison Lemma to equations
(\ref{Varia}) and $u''=0$. Then $u(x;C)\geq 0$ for $C\in[C_+^0,C_+^1]$
not only when $x>b$ but everywhere  in the interval where $u(x;C)$ is
defined. This implies that $\frac{\partial\psi_+(a;C)}{\partial C}\geq
0$ for $C\in[C_+^0,C_+^1]$ i.e. $\psi_+(a;C)$ is nondecreasing functon
of $C$. Therefore $\psi_+(a;C_+^0)\leq \psi_+(a;C_+^1)$ and this is a
contradiction to the assumption that one of the conditions
(\ref{Lemma1_1})-(\ref{Lemma1_2}) breaks at $C=C_+^1$. $\blacksquare$
\medskip

{\it Proof of Lemma 2}. Let us apply Comparison Lemma taking
\begin{eqnarray}
&&f(t,\xi)=-\Omega\xi+\xi^3;\quad g(t,\xi)=-(\omega-V(t))\xi+\xi^3\\
&&A=\sqrt{\Omega};\quad B=+\infty
\end{eqnarray}
It is straightforward to check that any solution of the equation
$\psi''+\Omega\psi-\psi^3=0$ with initial data
$\psi(t_0)\geq\sqrt{\Omega}$, $\psi'(t_0)$ -- arbitrary, is singular.
Therefore any solution of Eq(\ref{MainEqRep}) with the same initial
conditions is also singular. Due to oddness of the function $g(t,\xi)$
with respect to $\xi$ one concludes that any solution of
Eq(\ref{MainEqRep}) such that $\psi(t_0)\leq-\sqrt{\Omega}$,
$\psi'(t_0)$- arbitrary, also is singular. This implies the assertion
of Lemma 2. $\blacksquare$

\end{document}